\documentclass[twocolumn,english,showpacs,preprintnumbers,amsmath,amssymb]{revtex4}
\usepackage{pslatex}
\usepackage[T1]{fontenc}
\usepackage[latin1]{inputenc}
\usepackage{geometry}
\usepackage{amsmath}
\usepackage{graphicx}
\usepackage{amssymb}

\makeatletter
\usepackage{graphicx}
\usepackage{dcolumn}
\usepackage{bm}
\usepackage{epstopdf}

\usepackage{babel}
\makeatother
\begin{document}

\title{Photoassociation and optical Feshbach resonances in an atomic Bose-Einstein
condensate: treatment of correlation effects}

\author{Pascal Naidon}

\email{pascal.naidon@lac.u-psud.fr}

\author{Fran\c{c}oise Masnou-Seeuws}

\affiliation{Laboratoire Aim\'{e} Cotton, CNRS, B\^{a}t. 505 Campus d'Orsay,
91405 Orsay Cedex, France.}

\begin{abstract}
In this paper we formulate the time-dependent many-body theory of
photoassociation in an atomic Bose-Einstein condensate with realistic
interatomic interactions, using and comparing two approximations:
the first-order cumulant approximation, originally developed by K\"{o}hler
and Burnett {[}Phys. Rev. A \textbf{65}, 033601 (2002){]}, and the
reduced pair wave approximation, based on a previous paper {[}Phys.
Rev. A \textbf{68} 033612 (2003){]} generalizing to two channels the
Cherny-Shanenko approach {[}Phys. Rev. E \textbf{62}, 1046 (2000){]}.
The two approximations differ only by the way a pair of condensate
atoms is influenced by the mean field at short interatomic separations.
For these approximations we identify two different regimes of photoassociation:
the adiabatic regime and the coherent regime. The threshold for the
so-called {}``rogue dissociation'' {[}Phys. Rev. Lett. \textbf{88},
090403 (2002){]} (where mean-field theory breaks down) is found to
be different in each regime, which sheds new light on the experiment
of McKenzie \textit{et al} {[}Phys. Rev. Lett. \textbf{88}, 120403
(2002){]} and previous theoretical calculations.

We then use the two approximations to investigate numerically the
effects of rogue dissociation in a sodium condensate under conditions
similar to the McKenzie \textit{et al} experiment. We find two different
effects: reduction of the photoassociation rate at short times, and
creation of correlated pairs of atoms, confirming previous works.
We also observe that the photoassociation line shapes become asymmetric
in the first-order cumulant approximation, while they remain symmetric
in the reduced pair wave approximation, giving the possibility to
experimentally distinguish between the two approximations. 
\end{abstract}
\maketitle

\section{INTRODUCTION}

\label{sec:Intro} The possibility to create molecular condensates
opens new research avenues. These include test of fundamental symmetries
\cite{sauer1994,hinds1997}, determination of fundamental constants
through molecular spectroscopy with unprecedented accuracy, creation
of molecular lasers and invention of a coherent super-chemistry \cite{heinzen2000}
at ultralow temperatures. Since direct laser cooling or sympathetic
cooling appear more difficult for molecules than for atoms, many experimental
groups have worked on procedures to transform an atomic quantum degenerate
gas into a molecular condensate. Over the last two years a wealth
of new experimental results have appeared about formation of molecules
in a degenerate gas, starting either from an atomic Bose-Einstein
condensate \cite{donley2002,claussen2002,herbig2003,xu2003,durr2004}
or from an atomic Fermi gas \cite{regal2003,zwierlein2003,strecker2003,cubizolles2003,jochim2003,jochim2003a,regal2004}.\\

So far the most successful scheme to produce molecules in a condensate
involves an adiabatic sweep near a Feshbach resonance by varying in
time the strength of an external magnetic field. However, the resulting
Feshbach molecules are usually in a highly-excited vibrational level
of the ground state interatomic potential and decay rapidly due to
collisional quenching. An alternative would consist in varying the
frequency of a laser in a photoassociation experiment, thereby sweeping
across an optically induced Feshbach resonance. Such resonances have
already been discussed in theoretical papers \cite{fedichev1996,bohn1997,kokoouline2001}
and explored in recent experiments \cite{fatemi2000,theis2004,thalhammer2005}.
These resonances are in many ways similar to the magnetic Feshbach
resonances, except that the resonant state is an electronically excited
state with a usually very short lifetime due to spontaneous emission.
As we shall see in this paper, this qualitatively changes the many-body
properties of the system. On the other hand, photoassociation offers
more experimental possibilities by varying in time the frequency or/and
the intensity of the laser \cite{koch2005}. In particular, the use
of shaped laser pulses opens the way to better control in this domain.
For instance, by using a series of shaped pulses it should be possible
to quickly transfer vibrationally excited molecules created via a
magnetic Feshbach resonance to their ground vibrational state~\cite{koch2004a}.
Photoassociation with pulsed lasers could therefore solve the problem
of the short lifetime of molecular condensates, in particular bosonic
dimers.\\

Mixed atomic and molecular condensates, formed by magnetic or optically
induced Feshbach resonances, are typically described in the mean-field
approximation, corresponding to two coupled Gross-Pitaevski\u{\i}
equations \cite{drummond1998,timmermans1999,heinzen2000}. This may
not be sufficient when correlations play a significant role. For instance,
they must be introduced in the theoretical models \cite{vardi2001,kokkelmans2002a,mackie2002,kohler2003a,drummond2004}
to reproduce the damping in the observed \cite{donley2002} oscillations
between the atomic and the molecular components of a condensate exposed
to a time-dependent magnetic field. 

In the case of an isolated resonance, it may be sufficient to describe
the microscopic quantum dynamics with effective interactions such
as contact or separable potentials, involving parameters fitted on
two-body calculations. However, in the more general case of photoassociation
with shaped laser pulses, many levels may be involved; several theoretical
studies \cite{vala2001,luc-koenig2004b,luc-koenig2004c} investigating
photoassociation with chirped pulses at a two-body level have shown
a great sensitivity to the details of the molecular potentials.

A general treatment of photoassociation in a Bose-Einstein condensate
should therefore be able to take both correlations and realistic interactions
into account. To achieve this goal, we will consider two methods: 

\begin{itemize}
\item In a series of papers, \cite{kohler2002,kohler2003,kohler2003a,goral2004}
the Oxford group has developed a method based on a truncation of the
expansion of correlation functions in terms of non-commutative cumulants
\cite{fricke1996}. The method to first order, hereafter referred
to as \emph{first-order cumulant approximation}, has been used so
far assuming a separable interatomic potential, which was sufficient
to successfully interpret magnetic Feshbach experiments~\cite{donley2002,kohler2003}.\\

\item In a previous paper \cite{naidon2003}, hereafter referred to as paper
I, we have revisited the treatment of photoassociation and Feshbach
resonances following another approach, hereafter referred to as the
\emph{effective pair wave approach}. This work generalizes the ideas
of A. Yu Cherny and A. A. Shanenko \cite{cherny2000,cherny2002} to
two coupled channels.\\

\end{itemize}
The present paper goes further by numerically solving the effective
pair wave equations as well as the cumulant equations with realistic
potentials. It is organized as follows: 

\begin{itemize}
\item In Section \ref{sec:Correlations} we discuss the limitations of contact
potentials and present how to treat realistic interactions using either
the \emph{first-order cumulant} or the \emph{reduced pair wave} approximations. 
\item In Section \ref{sec:2channel} we give the coupled equations for a
two-channel modelling of photoassociation in a condensate, using the
two approximations described in Sec.~\ref{sec:Correlations}.
\item In Section \ref{sec:RogueDissociation} we make a connection between
these equations and the usual mean-field approximation of Ref.~\cite{timmermans1999}
for an isolated resonance. This enables us to determine the different
regimes and conditions for which correlations play an important role. 
\item Section \ref{sec:numerical} describes the numerical methods used
to solve the coupled equations given in Sec. \ref{sec:2channel}. 
\item In Section \ref{sec:results} we present and discuss the results of
our numerical calculations for high-intensity photoassociation in
a sodium condensate, starting from the experimental conditions of
Ref.~\cite{mckenzie2002}. 
\item We conclude in Section \ref{sec:conclu}. 
\end{itemize}
In this paper we give scattering lengths in units of the Bohr radius
$a_{0}$ = 0.529177 10$^{-10}$m.

\section{Different methods to treat the condensate dynamics\label{sec:Correlations}}

\subsection{Methods with a contact potential }

In many theoretical treatments of the condensed Bose gas \cite{pitaevskii2003,leggett2001},
the interaction potential $U(r)$ for two atoms is replaced by the
effective contact potential\begin{equation}
U_{\delta}(\mathbf{r})=\frac{4\pi\hbar^{2}a}{m}\delta^{3}(\mathbf{r}),\label{eq:ContactInteraction}\end{equation}
 where $a$ is the scattering length of the potential $U(r)$ for
particles of mass $m$. The physical argument for this replacement
is that the detailed structure of the potential $U$ is not resolved
at the scale of the typical de Broglie wavelength associated with
very low collision energies. Although the mathematical form of the
contact potential (\ref{eq:ContactInteraction}) does not lead to
a well defined scattering problem, it can nevertheless give sensible
results within theories which treat the interaction perturbatively.
For example, in the two-body problem, the Born approximation for the
scattering length of the contact potential is equal to the scattering
length $a$ of the original potential\begin{equation}
a_{Born}[U_{\delta}]=\frac{m}{4\pi\hbar^{2}}\int\!\! d^{3}\mathbf{r}\; U_{\delta}(\mathbf{r})=a.\label{eq:aBornContact}\end{equation}

Similarly, in the many-body problem the Hartree-Fock approximation
for atoms interacting with a contact potential leads to the well-known
Gross-Pitaevski\u{\i} equation~\cite{pitaevskii2003,leggett2001}.
However, beyond these first-order approximations, ultraviolet divergences
arise due to the zero-range nature of the contact potential, and one
must use a regularized delta function \cite{lee1957a}, or renormalize
the coupling constant of the contact potential \cite{kokkelmans2002}.
Once these technical procedures are applied, effective contact potentials
can be successful in many situations. In particular we mention the
description of damped oscillations in a mixed condensate \cite{kokkelmans2002a,mackie2002}. 

However, since the effective contact potentials eliminate the details
of the real potential, some physical quantities, such as the kinetic
and interaction energies of the gas \cite{cherny2002} cannot be predicted.
Similarly, the presence of bound levels in the potential is not directly
taken into account, whereas these bound levels can play an important
role in the dynamics of photoassociation with chirped laser pulses
\cite{luc-koenig2004b,luc-koenig2004c}. For these strongly time-dependent
situations the details of the potentials will be important, since
their structure is explored over a wide range of internuclear separations.
We expect that effective contact potentials may be inadequate and
a realistic interaction potential should be used.

\subsection{Methods using a realistic potential\label{sub:RealisticPotentials}}

Realistic interaction potentials $U(r)$ may have a deep well and
a strong repulsive wall at short interatomic separations $r$. Because
of these features, perturbative treatments like the Born approximation
are inadequate. For instance, the {}``Born'' scattering length of
the potential $U$ \begin{equation}
a_{Born}[U]=\frac{m}{4\pi\hbar^{2}}\int\!\! d^{3}\mathbf{r}\; U(r)\label{eq:aBornu}\end{equation}
is a poor approximation of the actual scattering length $a$. In fact,
because of the repulsive wall, most potentials are singular at $r=0$
so they cannot be integrated over space. As a result, $a_{Born}$
diverges in principle. In order to obtain finite results we need to
start from the exact expression for the scattering length\begin{equation}
a[U]=\frac{m}{4\pi\hbar^{2}}\int\!\! d^{3}\mathbf{r}\; U(r)\varphi_{2B}(r)\label{eq:aExact}\end{equation}
where $\varphi_{2B}$ is the solution of the two-body Schr\"{o}dinger
equation at zero energy: $\left(-\frac{\hbar^{2}\nabla^{2}}{m}+U(r)\right)\varphi_{2B}(\mathbf{r})=0$
and is normalised to be asymptotically equal to 1. The deviations
from 1 of the wave function $\varphi_{2B}$ represent the correlations
induced by the potential. In particular, because of the strong repulsive
wall, $\varphi_{2B}(r)$ vanishes for $r$ less than a few Bohr radii
as it is nearly impossible to find two atoms in this region. As a
result, the integrand $U(r)\varphi_{2B}(r)$ in Eq.~(\ref{eq:aExact})
is regular for short $r$, which leads to a finite $a[U]$. Interestingly,
Eq.~(\ref{eq:aExact}) can be seen as the perturbative expression
(\ref{eq:aBornu}) where the {}``bare'' potential is replaced by
a new effective potential $\mathcal{U}_{2B}(r)=U(r)\varphi_{2B}(r)$.

In a well-defined many-body theory we also expect that when the potential
appears in an equation it is always multiplied by many-body correlation
functions $\varphi_{MB}$ which go to zero for short interatomic separations
and are equal to 1 at large separations\cite{jastrow1955,yukalov1990,cherny2000}.
This introduces effective in-medium potentials $\mathcal{U}_{MB}(r)=U(r)\varphi_{MB}(r)$.
At low energy, most properties should be determined by the zero-momentum
Fourier component of $\mathcal{U}_{MB}$ given by the expression\begin{equation}
\frac{m}{4\pi\hbar^{2}}\int\!\! d^{3}\mathbf{r}\; U(r)\varphi_{MB}(r)\label{eq:ZeroMomentumEffectivePotential}\end{equation}

In the region where $U(r)$ is not negligible, many-body effects are
expected to be small and $\varphi_{MB}$ should be close to the two-body
wave function $\varphi_{2B}$, so that expression (\ref{eq:ZeroMomentumEffectivePotential})
is approximately equal to the scattering length $a[U]$ of the two-body
theory. This is consistent with the idea that the two-body scattering
length determines low-energy properties.

By analogy with the two-body theory, we may also expect that a well-defined
many-body theory for realistic interactions will have essentially
the same structure as a perturbative theory with the replacement $U(r)\rightarrow\mathcal{U}_{MB}(r)$.
This similarity was illustrated in Refs.~\cite{cherny2000,leggett2003},
where non-perturbative Bogoliubov equations were derived and an effective
interaction $\mathcal{U}_{MB}$ appeared in place of $U$. This replacement
$U(r)\rightarrow\mathcal{U}_{MB}(r)$ in the perturbative theory is
equivalent to the contact potential replacement $U(r)\rightarrow U_{\delta}(r)$
as long as we are only concerned with low-momentum quantities, such
as (\ref{eq:ZeroMomentumEffectivePotential}). However, we should
note that in general these replacements are not equivalent. Firstly,
the Fourier transforms of the effective in-medium potentials $\mathcal{U}_{MB}(r)$
are momentum dependent, whereas for contact potentials $U_{\delta}$
they are constant. Differences might therefore arise for high-momenta,
\emph{i.e.} short distances, the most important one being the absence
of ultraviolet divergences. Secondly, the potentials $\mathcal{U}_{MB}$
depend on the many-body dynamics through $\varphi_{MB}$, whereas
the contact potentials $U_{\delta}$ are solely determined by two-body
parameters. In particular, in the time-dependent case, the correlation
function is time-dependent and complex-valued, leading to a \emph{time-dependent
complex-valued} effective interaction $\mathcal{U}_{MB}(r,t)$. This
feature is not present for the contact interaction $U_{\delta}$,
which remains constant and real. This is another indication that in
time-dependent cases, models based on a contact interaction or an
explicit description of the short-range interaction might differ.
In any case the time-dependent structure of short-range correlations
in Bose-Einstein condensates is mostly unknown and deserves further
investigation.

\subsection{General theory with realistic potentials\label{sub:GeneralEquations}}

Let us first consider a Bose-condensed atomic gas with only one scattering
channel. The dynamics of this many-body system is given in the second-quantization
formalism by the Heisenberg equation of motion \begin{equation}
i\hbar\frac{\partial\hat{\psi}}{\partial t}(\textbf{x},t)=H_{\textbf{x}}\hat{\psi}(\textbf{x},\! t)+\int\!\! d^{3}\mathbf{z}~\hat{\psi}^{\dagger}(\textbf{z},t)U(|\textbf{z-x}|)\hat{\psi}(\textbf{z},t)\hat{\psi}(\mathbf{x},\! t),\label{eq:HeisenbergEquation}\end{equation}
where $\hat{\psi}(\mathbf{x})$ is the field operator annihilating
an atom at point $\mathbf{x}$, $H_{\textbf{x}}$ the one-body Hamiltonian
describing the motion of atoms in an external potential $V(\mathbf{x})$,
\begin{equation}
H_{\textbf{x}}=-\frac{\hbar^{2}\nabla_{\mathbf{x}}^{2}}{2m}+V(\mathbf{x})\label{eq:hamiltrap}\end{equation}
 and $U$ is the single-channel interaction potential between two
atoms (because of the very low density, we neglect interactiing potentials
involving more than two atoms). In the U(1) symmetry-breaking picture
\cite{leggett2001}, the condensate wave function $\Psi$ is given
by the quantum average of the field operator\begin{equation}
\Psi(\mathbf{x},t)=\langle\hat{\psi}(\mathbf{x},\! t)\rangle.\label{eq:DefinitionPsi}\end{equation}

The exact equation for $\Psi$ is therefore:

\begin{equation}
i\hbar\frac{\partial\Psi}{\partial t}(\textbf{x},t)=H_{\textbf{x}}\Psi(\textbf{x},\! t)+\int\!\! d^{3}\mathbf{z}~U(|\textbf{z-x}|)\langle\hat{\psi}^{\dagger}(\textbf{z},t)\hat{\psi}(\textbf{z},t)\hat{\psi}(\mathbf{x},\! t)\rangle\label{eq:ExactCondensateEquation}\end{equation}
According to Eq.~(\ref{eq:SimplifiedExpr1}) in the Appendix, assuming
that we may neglect the influence of the non-condensate atoms colliding
with the condensate atoms, we can write \begin{equation}
\langle\hat{\psi}^{\dagger}(\textbf{z},\! t)\hat{\psi}(\textbf{z},\! t)\hat{\psi}(\mathbf{x},\! t)\rangle\approx\Psi^{*}(\mathbf{z},\! t)\Phi(\mathbf{z,x},\! t),\label{eq:ApprOneBodyCollisionTerm}\end{equation}
 where we introduce the quantum average\[
\Phi(\mathbf{x,y},\! t)=\langle\hat{\psi}(\textbf{x},\! t)\hat{\psi}(\mathbf{y},\! t)\rangle.\]

This anomalous average is related to the pair wave function of two
condensate atoms (see Appendix). From Eq.~(\ref{eq:HeisenbergEquation}-\ref{eq:ApprOneBodyCollisionTerm}),
we get: \begin{equation}
i\hbar\frac{\partial\Psi}{\partial t}(\textbf{x},\! t)=H_{\textbf{x}}\Psi(\textbf{x},\! t)+\int\!\! d^{3}\mathbf{z}\Psi^{*}(\mathbf{z},\! t)U(|\textbf{z-x}|)\Phi(\mathbf{z,x},\! t),\label{eq:CondensateEquation}\end{equation}

This is a one-body Schr\"{o}dinger equation with an extra term corresponding
to the influence of surrounding condensate atoms. For convenience,
we can express the pair wave function $\Phi$ in terms of a product
of independent particle wave functions, 

\begin{equation}
\Phi(\mathbf{x,y},t)=\Psi(\mathbf{x},t)\Psi(\mathbf{y},t)\varphi(\mathbf{x,y},t),\end{equation}
so that all the correlation between two condensate atoms is concentrated
into the \emph{reduced pair wave function} $\varphi$ \cite{naidon2003,cherny2004}.
Assuming that $\Phi(\mathbf{x,y},t)$ is uncorrelated at large distances
$|\mathbf{x}-\mathbf{y}|$, we must have $\varphi(\mathbf{x,y},t)\approx1$
at such distances. Then we can write Eq. (\ref{eq:CondensateEquation})
as a Gross-Pitaevski\u{\i} equation \[
i\hbar\frac{\partial\Psi}{\partial t}(\textbf{x},\! t)=\Big(H_{\textbf{x}}+\frac{4\pi\hbar^{2}a_{M}(\mathbf{x},\! t)}{m}|\Psi(\mathbf{x},\! t)|^{2}\Big)\Psi(\textbf{x},\! t)\]
 by introducing a time and position dependent scattering length $a_{M}(\mathbf{x},t)$,
hereafter called the mean-field scattering length, \begin{equation}
a_{M}(\mathbf{x},t)=\frac{m}{4\pi\hbar^{2}}\int\!\! d^{3}\mathbf{z}~U(|\textbf{z-x}|)\varphi(\mathbf{z,x},t)\left|\frac{\Psi(\mathbf{z},\! t)}{\Psi(\mathbf{x},\! t)}\right|^{2}.\label{eq:MeanFieldScatteringLength}\end{equation}

We can see from this expression that the reduced pair wave function
$\varphi$ is the regularising correlation function $\varphi_{MB}$
discussed in Section \ref{sub:RealisticPotentials}. In the usual
Gross-Pitaevski\u{\i} theory, the pair wave function $\Phi$ is assumed
to be completely uncorrelated: $\varphi(\mathbf{x,y},t)\approx1$,
which leads to $a_{M}(\mathbf{x},t)\approx a_{Born}$, where $a_{Born}$
is the Born scattering length defined in Eq.~(\ref{eq:aBornu}).
If the true scattering length $a$ associated to the potential is
close to its Born approximation $a_{Born}$, then we retrieve the
standard Gross-Pitaevski\u{\i} equation for $\Psi$. However, we
have seen above that this is not the case for realistic molecular
potentials with a repulsive wall, since $a_{Born}(U)\rightarrow+\infty$.
We therefore have to keep the correlation $\varphi$ contained in
$\Phi$, so that its structure at short distance regularizes the singular
character of the potential $U$, leading to a value of the mean-field
scattering length $a_{M}$ close to the physical value $a$. 

The next step is to determine $\Phi(\mathbf{z,x},t)$. Starting from
the Heisenberg equation (\ref{eq:HeisenbergEquation}), one can derive
the exact equation for $\Phi$:\begin{multline}
i\hbar\frac{\partial\Phi}{\partial t}(\textbf{x,y},t)=(H_{\textbf{x}}+H_{\textbf{y}}+U(|\textbf{x-y}|))\Phi(\textbf{x,y},t)\\
+\left\{ \int\!\! d^{3}\mathbf{z}~U(\textbf{z-y})\langle\hat{\psi}^{\dagger}(\textbf{z},\! t)\hat{\psi}(\textbf{z},\! t)\hat{\psi}(\mathbf{x},\! t)\hat{\psi}(\mathbf{y},\! t)\rangle\right\} \\
+\{\mathbf{x}\leftrightarrow\mathbf{y}\}.\label{eq:ExactPairEquation}\end{multline}

Eq.~(\ref{eq:ExactPairEquation}) is a two-body Schr\"{o}dinger
equation with an extra term corresponding to the influence of the
medium. In order to take this term into account, we have to determine
the quantum average $\langle\hat{\psi}^{\dagger}(\textbf{z})\hat{\psi}(\textbf{z})\hat{\psi}(\mathbf{x})\hat{\psi}(\mathbf{y})\rangle$.
In fact, Eqs.~(\ref{eq:CondensateEquation}) and (\ref{eq:ExactPairEquation})
happen to be the first equations of an infinite set of equations coupling
each quantum average to higher-order quantum averages. Since we want
to truncate this set in order to restrict the dynamics to the condensate
wave function $\Psi$ and the pair wave function $\Phi$, we need
to approximate the average $\langle\hat{\psi}^{\dagger}(\textbf{z})\hat{\psi}(\textbf{z})\hat{\psi}(\mathbf{x})\hat{\psi}(\mathbf{y})\rangle$
in terms of $\Psi$ and $\Phi$.

\subsection{First-order cumulant approximation (FOC)}

\subsubsection{Truncation approximation}

A systematic method to truncate the infinite set of equations, named
the cumulant method, has been proposed by J. Fricke \cite{fricke1996}.
This method redefines the different quantum averages in terms of quantities
called non-commutative cumulants (see Appendix), which are constructed
in such a way that for a sufficiently weak interaction, they decrease
towards zero whith an increasing order. It is therefore natural to
neglect, in the infinite set of equations, all the cumulants of order
larger than a given order $n$. In this way, a consistent truncated
set of equations can be obtained for any desired order $n$. 

Doing so, one assumes that the system is close to the ideal gas. However,
this assumption might fail for realistic interactions which can induce
strong changes in the correlation structure at short distances. To
extend the validity of the cumulant method, T. K\"{o}hler and K.
Burnett \cite{kohler2002} have devised a modified version of the
truncation scheme where the free evolution of the cumulants of order
$n+1$ and $n+2$ is taken into account. This renormalizes the interaction
in the first $n$ equations, in a way similar to the regularization
we discussed in Section \ref{sub:RealisticPotentials}. As a consequence,
this extended cumulant approach is technically applicable to realistic
potentials, although it has been used so far only with effective separable
potentials in the context of atomic and molecular condensates \cite{goral2004,gasenzer2004}.
We show below how it can be applied to realistic potentials. 

Writing the cumulant equations to the first order, one finds Eqs.~(\ref{eq:CondensateEquation})
and (\ref{eq:ExactPairEquation}) with the following approximation:
\begin{equation}
\langle\hat{\psi}^{\dagger}(\textbf{z},t)\hat{\psi}(\textbf{z},t)\hat{\psi}(\mathbf{x})\hat{\psi}(\mathbf{y},t)\rangle\approx\Psi^{*}(\mathbf{z},t)\Phi(\mathbf{z,y},t)\Psi(\mathbf{x},t).\label{eq:FirstOrderCumulantApproximation}\end{equation}

The resulting coupled equations between the condensate wave function
and the pair wave function can be written as follows %
\footnote{Note that the cumulant equations are usually written with the correlated
part $\Phi(\mathbf{x,y})-\Psi(\mathbf{x})\Psi(\mathbf{y})$ of the
pair wave function, which corresponds to the second order cumulant
$\langle\hat{\psi}(\mathbf{x})\hat{\psi}(\mathbf{y})\rangle^{c}$.%
}

\begin{eqnarray}
i\hbar\frac{\partial\Psi}{\partial t}(\textbf{x},\! t) & = & \Big(H_{\textbf{x}}+M(\mathbf{x},\! t)\Big)\Psi(\textbf{x},\! t)\label{eq:FirstOrderCumulant1}\\
i\hbar\frac{\partial\Phi}{\partial t}(\textbf{x,y},\! t) & = & \Big(\! H_{\textbf{x}}+H_{\textbf{y}}+U(|\textbf{x-y}|)\Big)\Phi(\textbf{x,y},\! t)+\nonumber \\
~ & ~ & \Big(\! M(\mathbf{x},\! t)\!+\! M(\mathbf{y},\! t)\!\Big)\Psi(\textbf{x},\! t)\Psi(\textbf{y},\! t)\label{eq:FirstOrderCumulant2}\end{eqnarray}
involving the mean-field function \begin{equation}
M(\mathbf{x},t)=\frac{4\pi\hbar^{2}a_{M}(\mathbf{x},t)}{m}|\Psi(\mathbf{x},t)|^{2}.\label{eq:meanfield}\end{equation}
 These equations have been recently rederived from pair wave function
considerations \cite{cherny2004}. We notice that the mean field $M$
acts as an extra potential in Eq.~(\ref{eq:FirstOrderCumulant1})
but as a source term in Eq.~(\ref{eq:FirstOrderCumulant2}). We can
verify that in Eqs.~(\ref{eq:FirstOrderCumulant1},\ref{eq:FirstOrderCumulant2})
the interaction potential $U$ is indeed multiplied by the pair wave
function $\Phi$, which vanishes at short distances if the correct
boundary conditions are ensured: these equations can therefore be
used with a realistic molecular potential.

\subsubsection{Stationary states\label{sub:FOC-stationary-states}}

The cumulant method is designed mainly to propagate the dynamical
equations during a limited period of time \cite{kohler2002}. It is
not well designed for stationary states, which would require that
the equations are valid for an infinite period of time. However, in
order to solve the time-dependent equations (\ref{eq:FirstOrderCumulant1},\ref{eq:FirstOrderCumulant2}),
an initial condition is required. Here, one cannot assume that the
pair wave function is initially uncorrelated, as is proposed in Ref.~\cite{goral2004}:
in the case of realistic interactions such a choice would lead to
divergence, as we explained in Section \ref{sub:GeneralEquations}.
If we start from equilibrium, the initial condition has to be chosen
as close as possible to a physical stationary state, typically given
by the stationary Gross-Pitaevski\u{\i} equation. 

Setting $\Psi(\mathbf{x},t)=\bar{\Psi}(\mathbf{x})e^{-i\mu t/\hbar}$,
where $\mu$ is identified with the chemical potential we have to
set $\Phi(\mathbf{x,y},t)=\bar{\Phi}(\mathbf{x,y})e^{-i2\mu t/\hbar}$
accordingly to obtain, from Eqs.~(\ref{eq:FirstOrderCumulant1},\ref{eq:FirstOrderCumulant2}),
stationary equations for $\bar{\Psi}$ and $\bar{\Phi}$. Introducing
$\bar{M}(\mathbf{x})=M(\mathbf{x},0)$, these equations read: \begin{eqnarray}
\mu\bar{\Psi}(\textbf{x}) & = & \Big(H_{\textbf{x}}+\bar{M}(\mathbf{x})\Big)\bar{\Psi}(\textbf{x})\label{eq:StationaryFirstOrder1}\\
2\mu\bar{\Phi}(\textbf{x,y}) & = & \Big(H_{\textbf{x}}+H_{\textbf{y}}+U(|\textbf{x-y}|)\Big)\bar{\Phi}(\textbf{x,y})\nonumber \\
 &  & +\Big(\bar{M}(\mathbf{x})+\bar{M}(\mathbf{y})\Big)\bar{\Psi}(\textbf{x})\bar{\Psi}(\textbf{y}),\label{eq:StationaryFirstOrder2}\end{eqnarray}
and, as explained in Ref.~\cite{cherny2004}, lead to the stationary
Gross-Pitaevski\u{\i} equation for $\mu\rightarrow0$. However, in
typical experiments, $\mu$ can be much larger than the energy level
spacings of the trapping potential $V(\mathbf{x})$. In the experiment
of Ref.~\cite{mckenzie2002} for instance, the trap frequency is
about $200$ Hz, while the chemical potential $\mu\approx\frac{4\pi\hbar^{2}a}{m}\rho_{c}$
(where $\rho_{c}$ is the density at the center of the trap) is about
6 kHz. For chemical potentials of this order of magnitude, there is
a problem with the solution of Eqs. (\ref{eq:StationaryFirstOrder1}-\ref{eq:StationaryFirstOrder2}).
Indeed, when Eq.~(\ref{eq:StationaryFirstOrder1}) is solved formally
using the Green's function $G_{\mu}(\mathbf{x^{\prime},y^{\prime},x,y})$
of the operator $H_{\textbf{x}}+H_{\textbf{y}}+U(|\mathbf{x-y}|)-2\mu$:
\begin{multline}
\bar{\Phi}(\textbf{x,y})=\int d^{3}\mathbf{x^{\prime}}d^{3}\mathbf{y^{\prime}}G_{\mu}(\mathbf{x^{\prime},y^{\prime},x,y})\\
\Big(M(\mathbf{x^{\prime}})+M(\mathbf{y}^{\prime})\Big)\bar{\Psi}(\textbf{x}^{\prime})\bar{\Psi}(\textbf{y}^{\prime}),\label{eq:StationarySolutionFirstOrder}\end{multline}
 a singularity occurs in this Green's function each time $\mu$ is
close to an energy level of the trap. When Eqs.~(\ref{eq:StationaryFirstOrder1},\ref{eq:StationaryFirstOrder2})
and (\ref{eq:StationarySolutionFirstOrder}) are solved consistently,
this leads to series of unphysical resonances of the mean-field scattering
length as $\mu$ sweeps the series of energy levels of the trap (or
equivalently, as the density of the system is varied). As noted in
Ref.~\cite{cherny2004}, this problem can be fixed in the limit of
a homogeneous system: the Green's function becomes undefined for any
value of $\mu$, since singularities form a continuous spectrum in
this limit, and one can regularize it either by taking the principal
value of the integral in Eq.~(\ref{eq:StationarySolutionFirstOrder})
defining the formal solution, or by evaluating it slightly outside
the singularity line at $\mu\pm i\varepsilon$. The difficulty remains
however in a numerical implementation, since the calculations have
to be implemented within a limited box (see Section \ref{sec:numerical}),
leading to a discretization of the continuum. In between the spurious
resonances, the mean-field scattering length is equal to the correct
scattering length $a$, as we shall show in Section \ref{sec:numerical}.

\subsubsection{Conservation laws}

Although we have neglected the influence of the non-condensate atoms
in Eqs.~(\ref{eq:FirstOrderCumulant1}-\ref{eq:FirstOrderCumulant2}),
these atoms are present in the system and described by the second-order
cumulant 

\[
R'(\mathbf{x,y},t)=\langle\hat{\psi}^{\dagger}(\mathbf{x},t)\hat{\psi}(\mathbf{y},t)\rangle-\Psi^{*}(\mathbf{x},t)\Psi(\mathbf{y},t),\]
which we shall call the non-condensate density matrix. To first order
in the cumulant truncation scheme, this density matrix obeys the following
equation: \begin{multline}
i\hbar\frac{dR^{\prime}}{dt}(\textbf{x,y},t)\;=\;[H_{\textbf{y}}-H_{\textbf{x}}]R^{\prime}(\textbf{x,y},t)~+\\
\int\!\! d^{3}\textbf{z}\;\Phi^{\prime*}(\textbf{x,z},t)U(|\textbf{z-y}|)\Phi(\textbf{z,y},t)-\{\mathbf{x}\leftrightarrow\mathbf{y}\}^{*},\label{eq:FirstOrderCumulant3}\end{multline}
where $\Phi^{\prime}(\mathbf{x,y},t)=\Phi(\mathbf{x,y},t)-\Psi(\mathbf{x},t)\Psi(\mathbf{y},t)$
is the correlated part of the pair wave function. We can check \cite{goral2004}
that Eqs.~(\ref{eq:FirstOrderCumulant1}) and (\ref{eq:FirstOrderCumulant3})
lead to the conservation of the total number $N$ of particles, \begin{equation}
N=\int\!\! d^{3}\mathbf{x}\:|\Psi(\mathbf{x},t)|^{2}+\int\!\! d^{3}\mathbf{x}\; R^{\prime}(\mathbf{x,x},t).\label{eq:Conservation}\end{equation}

Moreover, one can deduce \cite{goral2004} the very interesting relation 

\begin{equation}
R'(\mathbf{x,y},t)=\int\!\! d^{3}\mathbf{z}\;\Phi^{\prime*}(\mathbf{x,z},t)\Phi^{\prime}(\mathbf{z,y},t),\label{eq:ApproximateBogoliubov}\end{equation}
which, in principle, guarantees the positivity of the non-condensate
density matrix, so that all occupation numbers of the non-condensate
modes remain positive. As noted in \cite{cherny2004}, this relation
is an approximation of the Bogoliubov relation \begin{multline}
R'(\mathbf{x,y},t)+\int\!\! d^{3}\mathbf{z}\; R^{\prime}(\mathbf{x,z},t)R^{\prime}(\mathbf{z,y},t)=\\
\int\!\! d^{3}\mathbf{z}\;\Phi^{\prime*}(\mathbf{x,z},t)\Phi^{\prime}(\mathbf{z,y},t)\end{multline}
 for small values of the condensate depletion. 

However, we should remark that the initial non-condensate fraction
predicted by Eq.~(\ref{eq:ApproximateBogoliubov}) is generally an
overestimation, because of the inappropriate long-range behaviour
of $\Phi$ within the first-order approximation. For instance, in
a homogeneous system, $\Phi\propto1-a/r$ at large distances $r$
(which is known to be incorrect from the Bogoliubov theory - see Ref.
\cite{lee1957a}), so that the right-hand side of Eq. (\ref{eq:ApproximateBogoliubov})
diverges. When calculations are performed within a box, the results
depend on the size of the box in an unphysical way. Therefore one
should discard the initial number of non-condensate atoms as meaningless,
considering only the subsequent variations. Note however that these
variations may become negative, which would be interpreted as negative
occupation numbers.

\subsection{Reduced pair wave approximation (RPW) \label{sub:Reduced-pair-wave} }

\subsubsection{Approximation on quantum averages}

We now start from another approach to close the dynamics of Eqs.~(\ref{eq:CondensateEquation})
and (\ref{eq:ExactPairEquation}). The pair wave function approach,
developed in paper I \cite{naidon2003}, and recalled in the Appendix,
shows that we can express some quantum averages in terms of eigenvectors
of the two-body density matrix, called pair wave functions. If we
retain only the contribution from the condensate atoms, (thus neglecting
that of the non-condensate atoms), we can express the quantum averages
of interest in terms of $\Psi$ and $\Phi$ exclusively. For instance,
the quantum average which appears in Eq.~(\ref{eq:ExactPairEquation})
can be approximated by \begin{multline}
\langle\hat{\psi}^{\dagger}(\textbf{z},t)\hat{\psi}(\textbf{z},t)\hat{\psi}(\mathbf{x},t)\hat{\psi}(\mathbf{y},t)\rangle\approx\\
\Psi^{*}(\mathbf{z},t)\frac{\Phi(\mathbf{z,x},t)\Phi(\mathbf{z,y},t)\Phi(\mathbf{x,y},t)}{\Psi(\mathbf{x},t)\Psi(\mathbf{y},t)\Psi(\mathbf{z},t)}.\label{eq:PairApproximation}\end{multline}

Compared to Eq.~(\ref{eq:FirstOrderCumulantApproximation}), this
approximation contains extra correlations between the coordinates,
in a way which respects the symmetry of the quantum average by permutation
of the last three coordinates. Note that in Eq.~(\ref{eq:ExactPairEquation})
this average is multiplied by the interaction potential $U(|\mathbf{z-y}|)$.
Moreover, since Eq.~(\ref{eq:ExactPairEquation}) is the equation
of motion for the pair wave function $\Phi(\mathbf{x,y})$, we suspect
that the correlation between the positions $\mathbf{z}$ and $\mathbf{y}$,
or between the positions $\mathbf{x}$ and $\mathbf{y}$, are the
most important ones. Neglecting the correlation contained in $\Phi(\mathbf{z,\! x})$,
we get the expression \begin{multline}
\langle\hat{\psi}^{\dagger}(\textbf{z},t)\hat{\psi}(\textbf{z},t)\hat{\psi}(\mathbf{x},t)\hat{\psi}(\mathbf{y},t)\rangle\approx\\
\frac{\Psi^{*}(\mathbf{z},t)}{\Psi(\mathbf{y},t)}\Phi(\mathbf{z,y},t)\Phi(\mathbf{x,y},t),\label{eq:ReducedPairApproximation}\end{multline}
 which now breaks the symmetry. 

The resulting equations can then be written \begin{eqnarray}
i\hbar\frac{\partial\Psi}{\partial t}(\textbf{x},t) & = & \Big(H_{\textbf{x}}+M(\mathbf{x},t)\Big)\Psi(\textbf{x},t),\label{eq:ReducedPairWave1}\\
i\hbar\frac{\partial\Phi}{\partial t}(\textbf{x,y},t) & = & \Big(H_{\textbf{x}}+H_{\textbf{y}}+U(\mathbf{x-y})+\nonumber \\
~ & ~ & M(\mathbf{x},t)+M(\mathbf{y},t)\Big)\Phi(\textbf{x,y},t),\label{eq:ReducedPairWave2}\end{eqnarray}
where the mean field $M(\textbf{x},t)$, defined as in Eq.~(\ref{eq:meanfield}),
now plays the role of a potential in both equations. Comparing with
Eq.~(\ref{eq:FirstOrderCumulant2}), one sees that a pair of condensate
atoms ``feels'' the mean field as a potential not only at large interatomic
distances, but also at shorter distances where the two atoms are correlated.
This is the only difference with the first-order cumulant equations
(\ref{eq:FirstOrderCumulant1}-\ref{eq:FirstOrderCumulant2}): although
it may seem minimal, we shall see in Section \ref{sec:results} that
it may have noticeable consequences on the dynamics. 

Using Eq.~(\ref{eq:ExactPairEquation}) and the approximation (\ref{eq:ReducedPairApproximation}),
we find the equation for $\varphi$: \begin{equation}
i\hbar\frac{\partial\varphi}{\partial t}(\mathbf{x,y},t)=\Big[-\frac{\hbar^{2}}{2m}(\nabla_{y}^{2}+\nabla_{y}^{2})+\mathcal{C}_{y}+\mathcal{C}_{x}+U(|\textbf{x-y}|)\Big]\varphi(\mathbf{x,y},t),\label{eq:ReducedPairEquation}\end{equation}
with $\mathcal{C}_{y}=-\frac{\hbar^{2}}{m}\vec{\nabla}_{y}\ln\Psi(\mathbf{y})\cdot\vec{\nabla_{y}}$.
Since the condensate wave function $\Psi$ is expected to be almost
uniform in the domain of variation of $\varphi$, we can neglect the
crossed terms $\mathcal{C}_{y}$ and $\mathcal{C}_{x}$ in Eq.~(\ref{eq:ReducedPairEquation}).
This condition is of course exact in a uniform system, where the crossed
terms are stricly zero. In such conditions, there is no momentum transfer
between the wave function of the condensate and the reduced pair wave
function: Eq.~(\ref{eq:ReducedPairEquation}) is then simply the
two-body Schr\"{o}dinger equation in free space. In principle, this
equation should contain other terms describing the influence of the
medium \cite{naidon2003}: since they are neglected within the approximation
(\ref{eq:ReducedPairApproximation}), we call it the \emph{reduced
pair wave approximation}.

\subsubsection{Stationary states}

The stationary states are easily found. Setting again $\Psi(\mathbf{x},t)=\bar{\Psi}(\mathbf{x})e^{-i\mu t/\hbar}$
and $\Phi(\mathbf{x,y},t)=\bar{\Phi}(\mathbf{x,y})e^{-i2\mu t/\hbar}$
, which implies $\varphi(\mathbf{x,y},t)=\varphi(\mathbf{x,y},0)=\bar{\varphi}(\mathbf{x,y})$,
we find: \begin{eqnarray}
\mu\bar{\Psi}(\textbf{x}) & = & (H_{\textbf{x}}+\bar{M}(\mathbf{x}))\bar{\Psi}(\textbf{x}),\label{eq:StationaryReducedPair1}\\
0 & = & \Big[-\!\frac{\hbar^{2}}{2m}(\nabla_{y}^{2}+\nabla_{y}^{2})+U(|\textbf{x-y}|)\Big]\bar{\varphi}(\mathbf{x,y})\qquad\label{eq:StationaryReducedPair2}\end{eqnarray}

From (\ref{eq:StationaryReducedPair2}), we see that independently
of the value of the chemical potential $\mu$, $\bar{\varphi}(\mathbf{x,y})$
is the stationary solution $\varphi_{2B}(|\mathbf{x-y}|)$ of the
two-body scattering problem at zero energy, with the limiting condition
$\bar{\varphi}\rightarrow1$ at large distances. As a result, the
mean-field scattering length (\ref{eq:MeanFieldScatteringLength})
is given by: \begin{eqnarray}
\bar{a}_{M}(\mathbf{x}) & = & \frac{m}{4\pi\hbar^{2}}\int\!\! d^{3}\mathbf{z}~U(|\textbf{z-x}|)\varphi_{2B}(|\textbf{z-x}|)\frac{|\bar{\Psi}(\mathbf{z})|^{2}}{|\bar{\Psi}(\mathbf{x})|^{2}}\qquad\label{eq:amoyen}\\
 & \approx & \frac{m}{4\pi\hbar^{2}}\int\!\! d^{3}\mathbf{r}~U(r)\varphi_{2B}(r)\label{eq:asimple}\end{eqnarray}
where we assumed, as discussed in paper I \cite{naidon2003}, that
the condensate wave function does not vary significantly at the scale
of the potential range. According to Eq.~(\ref{eq:aExact}), the
right hand side of Eq.~(\ref{eq:asimple}) is the scattering length
$a$, so that Eq.~(\ref{eq:StationaryReducedPair1}) is nothing but
the usual stationary Gross-Pitaevski\u{\i} equation. To this respect,
the reduced pair wave approximation (\ref{eq:ReducedPairApproximation})
is free of the difficulty encountered with the first-order cumulant
approximation (\ref{eq:FirstOrderCumulantApproximation}). This improvement
is obtained because the truncation of the higher-order quantum averages
is less strict: the reduced pair wave approximation includes the extra
term $\left(M(\mathbf{x})+M(\mathbf{y})\right)\Phi^{\prime}(\mathbf{x,y})$
coming from the three-body wave function. Note that this term is also
present in its perturbative form $\int\!\! d^{3}\mathbf{z}~\left(U(|\textbf{z-x}|)+U(|\textbf{z-y}|)\right)\vert\Psi(\mathbf{z})\vert^{2}\Phi^{\prime}(\mathbf{x,y})$
in the Hartree-Fock-Bogoliubov approximation. This suggests that a
more consistent extension of the reduced pair wave approximation should
generalize all the terms of the Hartree-Fock-Bogoliubov equations
to nonpertubative expressions.

\subsubsection{Conservation laws}

Using Eqs.~(\ref{eq:HeisenbergEquation}), we can deduce the exact
equation of motion for the non-condensate density matrix: \begin{eqnarray*}
i\hbar\frac{dR^{\prime}}{dt}(\textbf{x,y},t)=[H_{\textbf{y}}-H_{\textbf{x}}]R^{\prime}(\textbf{x,y},t)~+\\
\bigg\{\int\!\! d^{3}\textbf{z}\; U(|\textbf{z-y}|)\Big(\langle\hat{\psi}^{\dagger}(\mathbf{x},\! t)\hat{\psi}^{\dagger}(\mathbf{z},\! t)\hat{\psi}(\mathbf{z},\! t)\hat{\psi}(\mathbf{y},\! t)\rangle\\
-\Psi^{*}(\mathbf{x},\! t)\langle\hat{\psi}^{\dagger}(\mathbf{z},\! t)\hat{\psi}(\mathbf{z},\! t)\hat{\psi}(\mathbf{y},\! t)\rangle\Big)\bigg\}-\{\mathbf{x}\leftrightarrow\mathbf{y}\}^{*}.\end{eqnarray*}

Retaining only the contribution from the condensate atoms, Eqs. (\ref{eq:SimplifiedExpr1}-\ref{eq:SimplifiedExpr2})
from the Appendix show that $R^{\prime}$ satisfies Eq.~(\ref{eq:FirstOrderCumulant3}),
so that the conservation equation (\ref{eq:Conservation}) still holds.
In contrast, the approximate Bogoliubov relation (\ref{eq:ApproximateBogoliubov})
is not fulfilled any longer. We have not been able so far to find
an alternative relation which would guarantee the positivity of the
non-condensate occupation numbers.

\section{Two-channel coupled equations for photoassociation \label{sec:2channel}}

\subsection{Coupled equations for two atoms with a photoassociation laser}

\begin{figure}
\hfill{}\includegraphics[%
  width=0.45\textwidth]{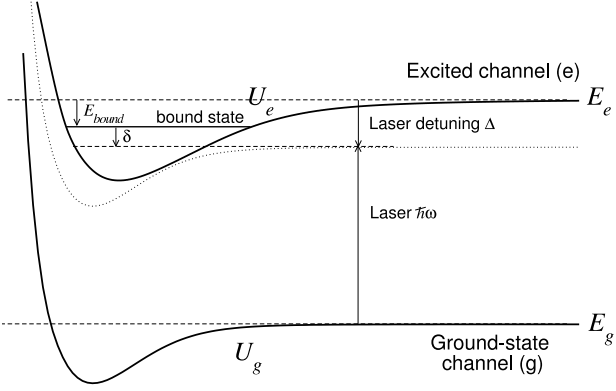}\hfill{}

\caption{\label{fig:pa}Scheme of the photoassociation process}
\end{figure}

Let us now turn to the standard two-channel model describing the photoassociation
reaction in a thermal gas of atoms, schematized in Fig.~\ref{fig:pa},
with a ground potential $U_{\text{g}}(r)$, an excited potential $U_{\text{e}}(r)+E_{e}$
and a time-dependent radiative coupling due to a laser red-detuned
by $\hbar\Delta$ relative to an atomic resonance line. We assume
that $U_{\text{e}}(\infty)\rightarrow E_{e}$, $U_{\text{g}}(\infty)\rightarrow0$,
$E_{e}$ being the dissociation limit of the excited potential, while
the origin of energy is the dissociation limit of the ground potential.
The relative motion of two atoms is described by a two-component wave
function $\Phi$, with $F_{\text{ground}}(\mathbf{r},t)$ describing
two atoms colliding in the potential $U_{\text{g}}$ (open channel),
and $F_{\text{exc}}(\mathbf{r},t)$ describing the ro-vibrational
motion in the excited potential $U_{\text{e}}$ (closed channel).
For a continuous laser with constant frequency $\omega$ such that
$\hbar\omega=\hbar\omega_{0}-\Delta$, in a classical model, the electric
field oscillates as $\cos(\omega t)$ and is coupled with the transition
dipole moment $\mathbf{D}(r)$ between the two electronic states.
At low temperature, only $s$-wave scattering should be considered,
and we shall neglect rotation effects and transfer of angular momentum
between the light field and the pair of atoms, so that instead of
($F_{\text{ground}}(\mathbf{r},t)$, $F_{\text{exc}}(\mathbf{r},t)$)
only radial functions $\phi_{\text{g}}(r,t)$, $\phi_{\text{exc}}(r,t)$
need to be introduced. Using the rotating wave approximation \cite{allen1987},
it is possible to eliminate the rapid oscillations in the coupling
term by defining new wave functions \begin{eqnarray}
\phi_{\text{e}}(r,t)=\exp(i\omega t)\phi_{\text{exc}}(r,t)\label{eq:rotating}\end{eqnarray}
 so that the the time-dependent Schr\"{o}dinger equation becomes
\cite{luc-koenig2004b}

\begin{eqnarray}
 &  & i\hbar\frac{\partial}{\partial t}\left(\!\begin{array}{c}
\phi_{g}(r,t)\\
\phi_{e}(r,t)\end{array}\!\right)\label{eq:2channelSchro}\\
 &  & =\left(\!\begin{array}{lc}
\hat{\mathbf{T}}+U_{g}(r) & W(r,t)\\
W(r,t) & \hat{\mathbf{T}}+U_{e}(r)+E_{e}-\hbar\omega\end{array}\!\right)\left(\!\begin{array}{c}
\phi_{g}(r,t)\\
\phi_{e}(r,t)\end{array}\!\right)\nonumber \end{eqnarray}

where we have introduced the kinetic energy operator $\hat{\mathbf{T}}$
and a coupling term: \begin{equation}
W(r,t)\approx W(t)=-\frac{1}{2}\sqrt{\frac{2I(t)}{c\epsilon_{0}}}D.\end{equation}
 $I(t)$ is the intensity of the laser, the constants $c$ and $\epsilon_{0}$
are respectively the ligth velocity and the vacuum permittivity, and
$D$ is one component of the dipole moment operator, depending upon
the polarisation of the laser, and assumed to be $r$-indedependent.
Since $E_{e}-\hbar\omega$=$\Delta$, the 2-channel Hamiltonian in
(\ref{eq:2channelSchro}) reads \begin{equation}
\mathbf{H}^{(2)}(r,t)=\left(\begin{array}{lc}
\hat{\mathbf{T}}+U_{g}(r) & W(t)\\
W(t) & \hat{\mathbf{T}}+U_{e}(r)+\Delta\end{array}\right)\end{equation}

The relevant quantities for the photoassociation reaction are illustrated
in Fig.~\ref{fig:pa}.

\subsection{Coupled equations for a condensate}

Considering now photoassociation in a condensate, we treat the many-body
problem with a two-component pair wave function $\Phi$: the usual
component $\Phi_{\text{g}}$ in the ground (open) channel, and a new
component $\Phi_{\text{e}}$ in the excited (closed) channel. The
component in the open channel is connected at large distances to a
product of condensate wave functions $\Psi$, whereas the component
in the closed channel corresponds to a bound vibrational level and
vanishes at large distances,\[
\Phi(\mathbf{x,y},\! t)=\left(\!\!\!\begin{array}{c}
\Phi_{\text{g}}(\mathbf{x,y},\! t)\\
\Phi_{\text{e}}(\mathbf{x,y},\! t)\end{array}\!\!\!\right)\xrightarrow[|\mathbf{x-y}|\rightarrow\infty]{\;}\left(\!\!\!\begin{array}{c}
\Psi(\mathbf{x},\! t)\Psi(\mathbf{y},\! t)\\
0\end{array}\!\!\!\right)\]

For simplicity, we will consider a homogeneous system. In this case,
the condensate wave function $\Psi(\mathbf{x},t)$ is uniform, and
the functions $\Phi(\mathbf{x,y},t)$ and $R^{\prime}(\mathbf{x,y},t)$
depend only on the relative coordinate $\mathbf{x-y}$ = $\mathbf{r}$.
Assuming an isotropic situation with s-wave scattering only we shall
simply write $\Psi(t)$, $\Phi_{g}(r,t)$, $\Phi_{e}(r,t)$ and $R^{\prime}(r,t)$
for the condensate wave function, the two components of the pair wave
function, and the density matrix for the non-condensate atoms in the
ground state.

\subsubsection{Coupled equations in the first-order cumulant approximation}

The first-order cumulant equations (\ref{eq:FirstOrderCumulant1}-\ref{eq:FirstOrderCumulant2})
are generalized as follows \cite{gasenzer2004}:\begin{eqnarray}
i\hbar\frac{\partial\Psi(t)}{\partial t} & = & M(t)\Psi(t)\label{eq:TwoChannelFirstOrder1}\\
i\hbar\frac{\partial}{\partial t}\left(\!\!\begin{array}{c}
\Phi_{\text{g}}(r,t)\\
\Phi_{\text{e}}(r,t)\end{array}\!\!\right) & = & \bigg(\mathbf{H}^{(2)}(r,t)\bigg)\!\cdot\!\left(\!\!\begin{array}{c}
\Phi_{\text{g}}(r,t)\\
\Phi_{\text{e}}(r,t)\end{array}\!\!\right)\nonumber \\
 &  & +\;2M(t)\left(\!\!\begin{array}{c}
\Psi^{2}(t)\\
0\end{array}\!\!\right)\label{eq:TwoChannelFirstOrder2}\end{eqnarray}

with the mean field: \begin{equation}
M(t)=\frac{1}{\Psi(t)}\int\!\! d^{3}\textbf{r}\Psi^{*}(t)\Big(U_{\text{g}}(r)\Phi_{\text{g}}(r,t)+W(r,t)\Phi_{\text{e}}(r,t)\Big)\label{eq:TwoChannelMeanField}\end{equation}
now depending upon the radiative coupling besides the ground state
potential, and upon both components of the pair wave function. From
$M(t)$ one can deduce a mean-field scattering length through \begin{equation}
a_{M}(t)=M(t)\frac{m}{4\pi\hbar^{2}|\Psi(t)|^{2}}.\label{eq:TwoChannelMeanScatt}\end{equation}
 The conservation equation reads \begin{equation}
\rho=\rho_{\text{g}}(t)+\rho_{\text{g}}^{\prime}(t)+\rho_{\text{e}}(t)\label{eq:DensityConservation}\end{equation}
where $\rho$ is the total density, $\rho_{\text{g}}(t)=|\Psi(t)|^{2}$
is the density of condensate atoms in the initial channel, $\rho_{\text{g}}^{\prime}(t)=R^{\prime}(0,t)$
is the density of non-condensate atoms in the initial channel, and
$\rho_{e}=4\pi\int\!\!|\Phi_{\text{e}}(r,t)|^{2}r^{2}dr$ is the density
of atoms in the excited (molecular) channel. Furthermore, we still
have in the ground channel the {}``approximate Bogoliubov relation\char`\"{}
(\ref{eq:ApproximateBogoliubov}), which now reads: \begin{equation}
R^{\prime}(r,t)=\int\!\! d^{3}\mathbf{z}\;\Phi_{\text{g}}^{\prime*}(\mathbf{z},t)\Phi_{\text{g}}^{\prime}(\mathbf{r-z},t).\label{eq:ApproximateBogoliubovRelation2}\end{equation}
and links the density matrix for the non-condensate atoms to the correlated
part of the component of the pair wave function in the ground channel.

\subsubsection{Coupled equations in the reduced pair wave approximation}

The generalized form of Eqs.~(\ref{eq:ReducedPairWave1}-\ref{eq:ReducedPairWave2})
is \begin{eqnarray}
i\hbar\frac{\partial\Psi(t)}{\partial t} & \!\!=\!\! & M(t)\Psi(t)\label{eq:TwoChannelReducedPair1}\\
i\hbar\frac{\partial}{\partial t}\left(\!\!\begin{array}{c}
\Phi_{\text{g}}(r,t)\\
\Phi_{\text{e}}(r,t)\end{array}\!\!\right) & \!\!=\!\! & \left(\!\mathbf{H}^{(2)}(r,t)+\!\left(\!\!\begin{array}{lc}
2M(t) & 0\\
0 & 0\end{array}\!\!\right)\!\right)\!\cdot\!\left(\!\!\begin{array}{c}
\Phi_{\text{g}}(r,t)\\
\Phi_{\text{e}}(r,t)\end{array}\!\!\!\right)\qquad\label{eq:TwoChannelReducedPair2}\end{eqnarray}
with the mean field $M(t)$ still defined by Eq.~(\ref{eq:TwoChannelMeanField}).
The conservation equation (\ref{eq:DensityConservation}) still holds,
but the Bogoliubov relation (\ref{eq:ApproximateBogoliubovRelation2})
is no longer valid.

\subsection{Introduction of the spontaneous emission}

\label{ssec:spont}

When a pair of atoms in a confining trap is photoassociated, populating
a bound vibrational level of the potential $U_{e}(r)$, the excited
molecule has a finite lifetime and decays back to the ground electronic
state by spontaneous emission of a photon. Most often, the final state
is a continuum level of the ground potential $U_{g}(r)$, where the
pair of atoms has enough energy to escape the trap. In some cases,
the radiative transition populates a bound level of $U_{g}(r)$, leading
to the formation of stable molecules \cite{fioretti1998,masnou-seeuws2001}.\\

In the previous equations, decay by spontaneous emission is not considered.
Assuming that it is mainly a loss phenomenon, it can be accounted
for by adding an imaginary term to the excited potential $U_{e}(r)$,
\begin{equation}
U_{e}(r)\rightarrow U_{e}(r)-i\frac{\hbar\gamma}{2},\label{eq:imagine}\end{equation}
where $2\pi/\gamma$ is the radiative lifetime of the bound levels
in the excited potential (we have assumed the dipole moment to be
$r$-independent). The component of the pair wave function in the
closed channel therefore contains an exponentially decreasing factor:
as a result, the total density (\ref{eq:DensityConservation}) is
no longer conserved during the time evolution.

\section{Mean-field equations and Rogue dissociation\label{sec:RogueDissociation}}

As a general rule, the coupled mean-field (Gross-Pitaevski\u{\i})
equations of Ref. \cite{timmermans1999} are retrieved from both approximations
when the pair dynamics can be eliminated adiabatically with respect
to the one-body dynamics \cite{naidon2003}. However the adiabaticity
condition given in Paper I \cite{naidon2003} (merely comparing the
coupling constants in the equations) has to be refined. We give here
a more relevant condition, related to the \emph{rogue dissociation}
analysis of Ref. \cite{javanainen2002}.

Let us first assume that only one excited molecular level is resonant.
In this case, we call $\delta$ the detuning of the laser from this
molecular level (see Fig. \ref{fig:pa}). One can write $\Phi_{e}(r,t)\approx\sqrt{2}\Psi_{m}(t)\varphi_{m}(r)$,
where $\varphi_{m}(r)$ is the volume-normalized wave function for
the relative motion in this resonant molecular level and $\Psi_{m}(t)$
is the (here uniform) centre-of-mass wave function corresponding to
a molecular condensate. With this normalization, $\vert\Psi_{m}\vert^{2}$
is the density of molecules.

Then, the crucial point is to write the correlation in the pair wave
function as a sum of an adiabatic correlation and a dynamic correlation\begin{eqnarray*}
\Phi_{g}(r,t) & = & \Psi^{2}(t)+\Phi_{g}^{ad}(r,t)+\Phi_{g}^{dyn}(r,t)\\
 & = & \Psi^{2}(t)\left(1+\varphi_{g}^{ad}(r,t)+\varphi_{g}^{dyn}(r,t)\right).\end{eqnarray*}

The adiabatic part is found by setting $\Phi_{g}^{dyn}$ and $\frac{\partial}{\partial t}\Phi_{g}^{ad}$
to zero in Eq. (\ref{eq:TwoChannelFirstOrder2}) or $\varphi_{g}^{dyn}$
and $\frac{\partial}{\partial t}\varphi_{g}^{ad}$ to zero in Eq.
(\ref{eq:TwoChannelReducedPair2}). We can then expand the correlation
in terms of the scattering states $\vert\varphi_{\textbf{k}}\rangle$
of the ground potential $U_{g}$ :\begin{eqnarray*}
\Phi_{g}(r,t) & = & \Psi^{2}(t)+\int\!\!\!\!\!\begin{array}{c}
\frac{d^{3}\textbf{k}}{(2\pi)^{3}}\end{array}\left(C_{\textbf{k}}^{ad}(t)+C_{\textbf{k}}^{dyn}(t)\right)\varphi_{\textbf{k}}(r)\end{eqnarray*}
with the normalization $\langle\varphi_{\textbf{k}}\vert\varphi_{\textbf{q}}\rangle=(2\pi)^{3}\delta^{3}(\textbf{k-q})$
\footnote{To be complete, one should also include the bound states of the potential.
This has no major consequence on the following analysis.%
}. We find:\begin{equation}
C_{\textbf{k}}^{ad}(t)=-\frac{m}{\hbar^{2}k^{2}}\left(g_{\textbf{k}}\Psi^{2}+w_{\textbf{k}}\Psi_{m}\right).\label{eq:AdiabaticCorrelation}\end{equation}
Eliminating this adiabatic part in Eqs. (\ref{eq:TwoChannelFirstOrder1}-\ref{eq:TwoChannelFirstOrder2})
or (\ref{eq:TwoChannelReducedPair1}-\ref{eq:TwoChannelFirstOrder2})
leads to the terms of the mean-field equations \cite{naidon2003}.
The equations for $\Psi$ and $\Psi_{m}$ then read:\begin{eqnarray}
i\hbar\dot{\Psi} & \!=\! & \Psi^{*}\left[\left(g_{0}\Psi^{2}+\int\!\!\!\!\!\begin{array}{c}
\frac{d^{3}\textbf{k}}{(2\pi)^{3}}\end{array}g_{\textbf{k}}C_{\textbf{k}}^{dyn}\right)+w\Psi_{m}\right]\label{eq:GrossPitaevskiiDyn1}\\
i\hbar\dot{\Psi}_{m} & \!=\! & \hbar(\delta^{\prime}\!-\! i\frac{\gamma}{2})\Psi_{m}+\!\frac{1}{2}\!\!\left(\! w_{0}\Psi^{2}\!+\!\!\int\!\!\!\!\!\begin{array}{c}
\frac{d^{3}\textbf{k}}{(2\pi)^{3}}\end{array}w_{\textbf{k}}C_{\textbf{k}}^{dyn}\!\right)\quad\label{eq:GrossPitaevskiiDyn2}\end{eqnarray}
where

\begin{itemize}
\item $g_{\textbf{k}}=\int d^{3}rU(r)\varphi_{\textbf{k}}(r)$ is the atom-atom
scattering coupling constant (in particular $g_{0}=\frac{4\pi\hbar^{2}a}{m}$)
\item $w_{\textbf{k}}=\sqrt{2}\langle\varphi_{\textbf{k}}\vert W\vert\varphi_{m}\rangle$
is the atom-molecule coupling constant 
\item $\delta^{\prime}=\delta+E_{self}$ is the detuning shifted by the
self-energy of the molecules
\end{itemize}
The equation for the dynamic correlation in the first-order cumulant
approximation is:\begin{equation}
i\hbar\dot{C}_{\textbf{k}}^{dyn}=\frac{\hbar^{2}\textbf{k}^{2}}{m}C_{\textbf{k}}^{dyn}-i\hbar\dot{C}_{\textbf{k}}^{ad}\label{eq:dynFOC}\end{equation}
and in the reduced pair wave approximation:

\begin{equation}
i\hbar\dot{c}_{\textbf{k}}^{dyn}=\frac{\hbar^{2}\textbf{k}^{2}}{m}c_{\textbf{k}}^{dyn}-i\hbar\dot{c}_{\textbf{k}}^{ad}\label{eq:dynRPW}\end{equation}
with $c_{\textbf{k}}=C_{\textbf{k}}/\Psi^{2}$. As long as the momentum
distribution of $C_{\textbf{k}}^{dyn}$ lies in the Wigner's threshold
regime, we can make the simplification: $g_{\textbf{k}}\approx g_{0}$
and $w_{\textbf{k}}\approx w_{0}$. Note that this approximation does
not lead to any ultraviolet divergence, because we have eliminated
the adiabatic correlation beforehand. Such an approximation, reminiscent
of the use of a contact potential, would have given rise to ultraviolet
divergences if we had made it in the original equations (\ref{eq:TwoChannelFirstOrder1}-\ref{eq:TwoChannelFirstOrder2})
or (\ref{eq:TwoChannelReducedPair1}-\ref{eq:TwoChannelFirstOrder2}).

It is now clear from Eqs. (\ref{eq:GrossPitaevskiiDyn1}-\ref{eq:GrossPitaevskiiDyn2}),
that when the dynamic correlation $C_{\textbf{k}}^{dyn}$ is negligible,
one gets the Gross-Pitaevski\u{\i} equations of Ref. \cite{timmermans1999}:\begin{eqnarray}
i\hbar\dot{\Psi} & = & g_{0}\vert\Psi\vert^{2}\Psi+w_{0}\Psi^{*}\Psi_{m}\label{eq:GrossPitaevskii1}\\
i\hbar\dot{\Psi}_{m} & = & \hbar(\delta^{\prime}-i\frac{\gamma}{2})\Psi_{m}+\frac{1}{2}w_{0}\Psi^{2}\label{eq:GrossPitaevskii2}\end{eqnarray}

The mean-field approximation will break down when the dynamic correlation
is not negligible any more. This means that molecules break into pairs
of atoms which will contribute to the non-condensate fraction instead
of the condensate fraction. From Eqs. (\ref{eq:GrossPitaevskiiDyn1}-\ref{eq:GrossPitaevskiiDyn2})
we see that such \emph{rogue dissociation} \cite{javanainen2002}
can be neglected as long as:

\begin{equation}
\left|\int\!\!\!\!\!\begin{array}{c}
\frac{d^{3}\textbf{k}}{(2\pi)^{3}}\end{array}C_{\textbf{k}}^{dyn}\right|\ll\left|\Psi^{2}\right|\label{eq:ConditionRogue}\end{equation}
To give a more explicit condition, we first have to make a distinction
between two different regimes of Eqs. (\ref{eq:GrossPitaevskii1}-\ref{eq:GrossPitaevskii2}).

\begin{figure*}
\hfill{}\includegraphics[%
  clip,
  width=0.90\textwidth]{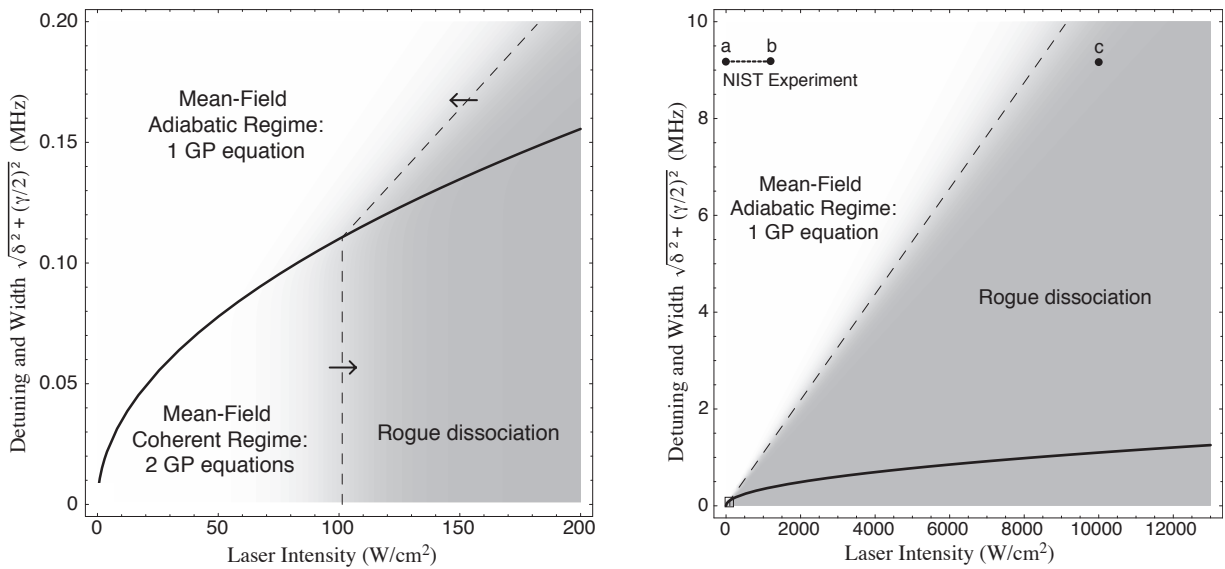}\hfill{}

\caption{{\small Different regimes of photoassociation in a condensate, as
a function of laser intensity and the combined effect of detuning
and spontaneous emission. These graphs correspond to the sodium resonance
described in \ref{ssec:expcond}, for a density of $4\times10^{14}$at/cm$^{3}$,
but spontaneous emission is taken as a free parameter. Left panel:
the solid line is the limit between the adiabatic regime and the coherent
regime of the mean-field equations (\ref{eq:GrossPitaevskii1}-\ref{eq:GrossPitaevskii2}).
The mean-field approximation is valid in the white areas, and rogue
dissociation occurs in the shaded areas. The vertical dashed lines
corresponds to the left hand side of (\ref{eq:ConditionRogue1}) being
equal to $\pi^{2}$ and the oblique dashed line corresponds to the
left hand side of (\ref{eq:ConditionRogue2}) being equal to $\pi^{2}$.
The arrows indicate how these lines move when the density is increased.
Right panel: larger view showing the experimental investigation of
Ref. \cite{mckenzie2002} (from dot} \emph{\small a} {\small to dot}
\emph{\small b}{\small ) where the spontaneous emission parameter
$\gamma/2$ is equal to $2\pi\times$9.18 MHz; dot} \emph{\small c}
{\small corresponds to the condition of our numerical investigation
described in \ref{ssec:results}. }}

\label{fig:rogue}
\end{figure*}

\subsection{The adiabatic regime}

We consider the limit when $\vert i\hbar\dot{\Psi}_{m}\vert\ll\vert\frac{1}{2}w_{0}\Psi^{2}\vert$.
In this case, $\Psi_{m}$ can be eliminated adiabatically, so that:

\begin{equation}
\Psi_{m}\approx-\frac{w_{0}}{2\hbar(\delta^{\prime}-i\frac{\gamma}{2})}\Psi^{2}\label{eq:AdiabaticBeta}\end{equation}
and:\begin{equation}
i\hbar\dot{\Psi}\approx\big(g-\frac{w_{0}^{2}}{2\hbar(\delta^{\prime}-i\frac{\gamma}{2})}\big)\vert\Psi\vert^{2}\Psi\equiv\frac{4\pi\hbar^{2}a_{M}}{m}\vert\Psi\vert^{2}\Psi\label{eq:AdiabaticAlpha}\end{equation}
 In this regime, the coupled equations reduce to a single Gross-Pitaevski\u{\i}
equation \cite{timmermans1999,mackie2002} where the mean-field scattering
length $a_{M}$ corresponds to the modified scattering length: \[
A=a+a_{\text{opt}}+ib_{\text{opt}}\]
 given by the two-body theory of the resonance \cite{fedichev1996,bohn1997}.
Therefore, most physical properties can be described by the usual
two-body theory. For instance, Eq. (\ref{eq:AdiabaticAlpha}) implies
that the atomic density $\rho_{g}=\vert\Psi\vert^{2}$ follows a simple
rate equation:\begin{equation}
\dot{\rho}_{g}=-\frac{8\pi\hbar b_{\text{opt}}}{m}\rho_{g}^{2}\label{eq:RateEquation}\end{equation}
and using (\ref{eq:AdiabaticCorrelation}) and (\ref{eq:AdiabaticBeta}),
one can calculate the pair wave function and find the usual asymptotic
behaviour:\begin{equation}
\Phi_{g}(r,t)\xrightarrow[r\to\infty]{}\Psi^{2}(t)\left(1-\frac{a_{M}}{r}\right)\label{eq:AsymptoticBehaviour}\end{equation}

As a matter of fact, the molecular field $\Psi_{m}$ scales as $\rho$
in this regime: it is merely a two-body field playing the role of
an intermediate state during the collision process. Using (\ref{eq:AdiabaticBeta})
and (\ref{eq:AdiabaticAlpha}), we find that the condition $\vert i\hbar\dot{\Psi}_{m}\vert\ll\vert\frac{1}{2}w_{0}\Psi^{2}\vert$
is satisfied whenever:\begin{equation}
\hbar\vert\delta^{\prime}+i\frac{\gamma}{2}\vert\gg w_{0}\sqrt{\rho},\;2g\rho\label{eq:ConditionAdiabatique}\end{equation}

This corresponds to the \emph{off-resonant regime} of Ref. \cite{timmermans1999}:
the detuning and the spontaneous emission are large with respect to
the coherent couplings.

\subsection{The coherent regime}

On the other hand, we may consider the limit $\vert\hbar(\delta^{\prime}-i\frac{\gamma}{2})\Psi_{m}\vert\ll\vert\frac{1}{2}w_{0}\Psi^{2}\vert$
and $\vert g_{0}\vert\Psi\vert^{2}\Psi\vert\ll\vert w_{0}\Psi^{*}\Psi_{m}\vert$.
In this case, the solutions of Eqs. (\ref{eq:GrossPitaevskii1}-\ref{eq:GrossPitaevskii2})
for an initially all atomic system are:\begin{eqnarray*}
\Psi & \approx & \sqrt{\rho}/\cosh(w_{0}\sqrt{\rho}t)\\
\Psi_{m} & \approx & -i\sqrt{\rho}\tanh(w_{0}\sqrt{\rho}t)\end{eqnarray*}

There is a coherent conversion between atoms and molecules. In this
regime, $\Psi_{m}$ scales as $\sqrt{\rho}$ and plays the role of
a one-body field, i.e. a molecular condensate in its own right. As
a result, the system cannot be described by a two-body theory any
more. For instance, the pair wave function is still of the form (\ref{eq:AsymptoticBehaviour}),
but the mean-field scattering length $a_{M}$ is now a time-dependent
quantity $a-iL\sinh(2w_{0}\sqrt{\rho}t)$ where:\[
L=\frac{m}{4\pi\hbar}\frac{w_{0}}{2\sqrt{\rho}}\]
is a many-body length which depends on the density. This means that
the density of atoms does not follow a rate equation.

We find that this coherent regime occurs when:\begin{equation}
w_{0}\sqrt{\rho}\gg\hbar\vert\delta^{\prime}+i\frac{\gamma}{2}\vert,\;2g\rho\label{eq:ConditionCoherence}\end{equation}

In practice, while it is possible to set the detuning to zero by varying
the laser frequency, the spontaneous decay $\gamma$ is fixed by the
molecular level of the system. Spontaneous emission in alkali systems,
and loss processes in general, make it very difficult to reach the
coherence condition (\ref{eq:ConditionCoherence}) with realistic
densities and laser intensites. For this reason, photoassociation
experiments in condensates so far \cite{mckenzie2002,theis2004,winkler2005}
have been confined to the adiabatic regime (\ref{eq:ConditionAdiabatique}),
and can be described simply by two-body theories. 

However, we made some estimates which indicate that photoassociation
for weakly-allowed transitions, such as those found in alkaline-earth
dimers, would lead to the coherent regime. The reason is that spontaneous
emission scales as the square of the dipole moment whereas the coherent
coupling $w_{0}\sqrt{\rho}$ only scales as the dipole moment.

\subsection{Rogue dissociation}

In the coherent regime, which is the regime originally investigated
in Ref. \cite{javanainen2002}, the relevant energy scale set by the
dynamics is $\hbar\Omega=w_{0}\sqrt{\rho}$ defining the coherent
Rabi frequency $\Omega/2\pi$. As a result, Eq. (\ref{eq:dynFOC})
or (\ref{eq:dynRPW}) can be made dimensionless using the characteristic
time $\tau=\Omega^{-1}$ and length $\xi=\sqrt{\hbar/m\Omega}=(4\pi L\rho)^{-1/2}$.
Since $\vert\Psi\vert^{2}\sim\rho$ and $\vert C_{\textbf{k}}^{dyn}\vert\sim1$,
the rogue dissociation condition (\ref{eq:ConditionRogue}) can be
written:\begin{equation}
\left(\rho\xi^{3}\right)^{-1}=\left(\frac{\hbar\Omega}{\frac{\hbar^{2}\rho^{2/3}}{m}}\right)^{3/2}=\left((4\pi L)^{3}\rho\right)^{1/2}\ll2\pi^{2}\label{eq:ConditionRogue1}\end{equation}

This condition was presented in J. Javanainen and M. Mackie's work
\cite{javanainen2002}.

Similarly, in the adiabatic regime, the relevant energy scale is $\hbar\Gamma=w_{0}^{2}\rho/\sqrt{\delta^{\prime2}+(\gamma/2)^{2}}$
corresponding to the mean field energy $\hbar\Gamma=\frac{4\pi\hbar^{2}\vert A\vert}{m}\rho$.
Eq. (\ref{eq:dynFOC}) or (\ref{eq:dynRPW}) can be made dimensionless
using the characteristic time $\tau_{0}=\Gamma^{-1}$ and length $\xi_{0}=\sqrt{\hbar/m\Gamma}=(4\pi\vert A\vert\rho)^{-1/2}$
(note that this is the usual healing length), and the rogue dissociation
condition is now:\begin{equation}
\left(\rho\xi_{0}^{3}\right)^{-1}=\left(\frac{\hbar\Gamma}{\frac{\hbar^{2}\rho^{2/3}}{m}}\right)^{3/2}=\left((4\pi\vert A\vert)^{3}\rho\right)^{1/2}\ll2\pi^{2}\label{eq:ConditionRogue2}\end{equation}

Both conditions (\ref{eq:ConditionRogue1}) and (\ref{eq:ConditionRogue2})
can be written as $\sqrt{\rho a_{M}^{3}}\ll1$ which is a straightforward
generalization of the usual condition $\sqrt{\rho a^{3}}\ll1$ for
the validity of the Gross-Pitaevski\u{\i} equation. This can be interpreted
as follows: rogue dissociation occurs when the average spacing $\rho^{-1/3}$
between the atoms becomes of the order of the typical length associated
to the adiabatic correlation, either $A$ in the adiabatic regime,
or $L$ in the coherent regime. However, there is a fondamental difference.
$A$ is the modified scattering length, a two-body quantity independent
of the density; in the adiabatic regime, one can thus reach the rogue
dissociation regime by increasing the density, so that the average
spacing between the atoms becomes of the order of $A$. The mean-field
equations are therefore valid at \emph{low density} in the adiabatic
regime. On the other hand, in the coherent regime, the many-body length
$L$ decreases with density, and more rapidly than the average spacing
between the atoms. As a result, one should decrease the density to
observe rogue dissociation. The mean-field equations are therefore
valid at \emph{high density} in the coherent regime. An overview of
the different regimes and conditions is given in Fig. \ref{fig:rogue}.

\section{Numerical methods and calculations}

\label{sec:numerical}

We now turn to the numerical resolution of the time-dependent equations
(\ref{eq:TwoChannelFirstOrder1}-\ref{eq:TwoChannelFirstOrder2})
and (\ref{eq:TwoChannelReducedPair1}-\ref{eq:TwoChannelReducedPair2}).
The main difficulty is that they describe simultaneously the microscopic
dynamics, with short characteristic times, and the macroscopic dynamics,
with longer ones. At each time step, the mean field $M(t)$ and the
mean-field scattering length $a_{M}(t)$ must be determined through
Eqs.~(\ref{eq:TwoChannelMeanField},\ref{eq:TwoChannelMeanScatt}),
from the two components of the pair wave function $\Phi_{\text{g}}(r,t)$
and $\Phi_{\text{e}}(r,t)$ and from the condensate wave function
$\Psi(t)$. Then the mean field influences the evolution of the three
wave functions in the coupled equations (\ref{eq:TwoChannelFirstOrder1},\ref{eq:TwoChannelFirstOrder2})
or (\ref{eq:TwoChannelReducedPair1},\ref{eq:TwoChannelReducedPair2}).
Note that in previous calculations based on the cumulant equations
\cite{gasenzer2004}, the use of a non-local, separable potential
made it possible to first perform an independent integration of the
two-body equation (\ref{eq:TwoChannelReducedPair2}), then inject
the results into the condensate equation to solve a non-Markovian
non-linear Schr\"{o}dinger equation. This simplifying procedure is
not implemented in the present paper where we use the same realistic
potentials in both approaches.\\

The pair wave functions are represented on a grid, using a mapping
procedure where the grid steps are adjusted to the local de Broglie
wavelengths \cite{kokoouline1999} in the two potentials, choosing
the smallest of the two values at a given $r$. The short range oscillations
are thus described with a dense grid and the long range behaviour
with a diffuse one. Typical value of the grid length $L$ is 200,000
$a_{0}$. The whole set of equations, either (\ref{eq:TwoChannelFirstOrder1}-\ref{eq:TwoChannelFirstOrder2})
or (\ref{eq:TwoChannelReducedPair1}-\ref{eq:TwoChannelReducedPair2}))
are then solved, using the standard Crank-Nicholson method \cite{press1996}
for the propagation in time. The solutions at $t=0$ are chosen as
stationary solutions for the open (ground) channel, when no laser
coupling is present.

\subsection{Initial state in the first-order cumulant approximation}

\begin{figure}
\hfill{}\includegraphics[%
  width=0.45\textwidth]{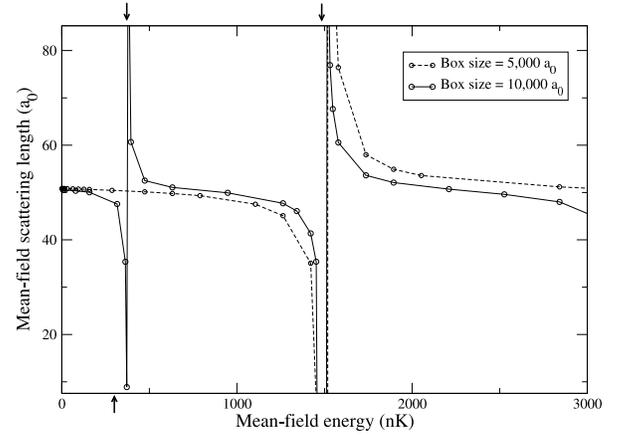}\hfill{}

\caption{{\small Mean-field scattering length as defined from Eqs.~(\ref{eq:TwoChannelMeanField}-\ref{eq:TwoChannelMeanScatt}),
calculated from the stationary solution of the first-order cumulant
equations (\ref{eq:FirstOrderCumulant1}, \ref{eq:FirstOrderCumulant2}),
as a function of the mean-field energy. The numerical calculations
shown here correspond to the case of a sodium condensate without any
photoassociation laser, and are performed within a square box of width
$d$= 10,000~a$_{0}$ (solid lines) or $d$=5,000~a$_{0}$ (dashed
line). The position of the first two levels of the wider box are indicated
by arrows on the upper horizontal axis, the second one coinciding
with the first level of the $d$=5,000 box. As explained in the text,
the mean-field scattering length has an unphysical divergence each
time the mean-field energy coincides with an energy level of the box.
The latter depend on the arbitrary size the box; in our calculations,
this size can typically go up to 200,000 atomic units, which increases
the number of resonances. The typical mean-field energy in the experiment
of Ref.~\cite{mckenzie2002} is about 6~kHz $\approx$ 300~nK,
and indicated by an arrow on the lower horizontal axis.}  }

\label{fig:ghost}
\end{figure}

Stationary solutions (in the rotating frame defined by Eq.~(\ref{eq:rotating}))
of Eqs. (\ref{eq:TwoChannelFirstOrder1}-\ref{eq:TwoChannelFirstOrder2})
are given by \begin{eqnarray}
\bar{\Psi} & = & \sqrt{\bar{\rho}_{\text{g}}}\label{eq:TwoChannelStationaryFirstOrder1}\\
\left(\!\!\begin{array}{c}
\bar{\Phi}_{\text{g}}(r,t)\\
\bar{\Phi}_{\text{e}}(r,t)\end{array}\!\!\right) & = & -\bigg(\!\mathbf{H}^{(2)}(r,t)\!-2\mu\bigg)^{-1}\!\cdot\!2\mu\left(\!\!\begin{array}{c}
\bar{\rho}_{\text{g}}\\
0\end{array}\!\!\right)\label{eq:TwoChannelStationaryFirstOrder2}\end{eqnarray}
where $\bar{\rho}_{\text{g}}$ is the initial condensate density,
and $\mu=\bar{M}$ is the chemical potential (equal to the initial
mean field in the present case). When the laser is initially off,
the coupling term $W$ in $\mathbf{H}^{(2)}$ is zero. To compute
(\ref{eq:TwoChannelStationaryFirstOrder1}-\ref{eq:TwoChannelStationaryFirstOrder2}),
we start from a given value for the condensate density $\bar{\rho}_{\text{g}}$
and a guess value for $\mu$. We determine the pair wave function
from (\ref{eq:TwoChannelStationaryFirstOrder2}) by performing a matrix
inversion. This gives a new value of the mean field through Eq.~(\ref{eq:TwoChannelMeanField}),
and the procedure is iterated until convergence is reached for $\mu$.
As explained in Section \ref{sec:Correlations}, for certain values
of the density $\bar{\rho}_{\text{g}}$, the matrix is singular and
the solution shows an unphysical resonance of the mean-field scattering
length. Between the resonances the mean-field scattering length is
essentially equal to the expected scattering length of the two-atom
system. This difficulty is illustrated in Fig.~\ref{fig:ghost}.
We therefore choose an initial density that is both close to the experimental
value and such that the mean-field scattering length does lie in-between
the resonances. 

\subsection{Initial state in the reduced pair wave approximation}

Stationary solutions of Eqs.~(\ref{eq:TwoChannelReducedPair1}-\ref{eq:TwoChannelReducedPair2})
are obtained by solving \begin{eqnarray}
\bar{\Psi} & = & \sqrt{\bar{\rho}_{g}}\label{eq:TwoChannelStationaryReducedPair1}\\
0 & = & \left(\!\mathbf{H}^{(2)}(r,t)\!-\!\left(\!\!\begin{array}{cc}
0 & 0\\
0 & 2\mu\end{array}\!\!\right)\right)\!\cdot\!\left(\!\!\begin{array}{c}
\bar{\Phi}_{\text{g}}(r,t)\\
\bar{\Phi}_{\text{e}}(r,t)\end{array}\!\!\right)\label{eq:TwoChannelStationaryReducedPair2}\end{eqnarray}

The two-component function $(\bar{\Phi}_{\text{g}},\bar{\Phi}_{\text{e}})$
is therefore the zero energy stationary solution of the two-channel
problem, the potential $U_{e}$ being shifted down by twice the chemical
potential. This means that the energy of the optically induced Feshbach
resonance is shifted by the mean-field energy of the condensate atoms,
even though the collision energy is nearly zero. To determine eigenfunctions
of Eq.~(\ref{eq:TwoChannelStationaryReducedPair2}) we propagate
Eq.~(\ref{eq:TwoChannelReducedPair2}) in imaginary time with the
Crank-Nicholson scheme \cite{press1996}. The finite value of the
time step value acts as an energy filter, which selects the zero energy
scattering state in a few steps.

\subsection{Accuracy check: computation of the scattering length}

\begin{figure}
\hfill{}\includegraphics[%
  width=0.45\textwidth]{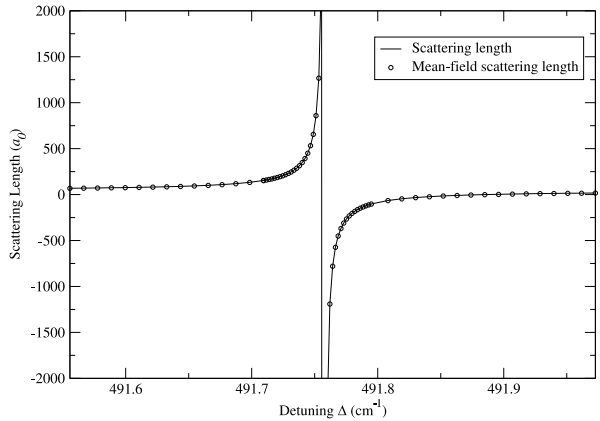}\hfill{}

\caption{{\small Optically induced Feshbach resonance: variation of the sodium
scattering length as a function of the detuning of the photoassociation
laser in the vicinity of the $v$=135, $J$=1 resonance \cite{mckenzie2002},
for a laser intensity of 15 kW/cm$^{2}$, and a variable detuning
$\Delta$ relative to the D$_{1}$ atomic resonance line. Spontaneous
emission is not included in these calculations. Solid line: scattering
length determined by phase shift from two-atom calculations. Dots:
mean-field scattering length as defined by Eqs.~(\ref{eq:TwoChannelMeanField}-\ref{eq:TwoChannelMeanScatt}),
calculated from the stationary states in the reduced pair wave approximation,
see Eq.~(\ref{eq:TwoChannelStationaryReducedPair1}-\ref{eq:TwoChannelStationaryReducedPair2}).}}

\label{fig:reson}
\end{figure}

Checks on the macroscopic dynamics can be performed by comparing with
the mean-field equations (\ref{eq:GrossPitaevskii1}-\ref{eq:GrossPitaevskii2})
when rogue dissociation is negligible (see below). As for the microscopic
dynamics, we have to ensure that the mean-field scattering length
(\ref{eq:TwoChannelMeanScatt}), deduced from the integral expression
(\ref{eq:TwoChannelMeanField}), is computed accurately enough at
each time step. To that purpose, we have solved the stationary version
of the two-body equations (\ref{eq:2channelSchro}) and extracted
the phase shift of the continuum wave function in the open channel
from its asymptotic behaviour. The scattering length can be deduced
by extrapolation to zero collision energy. In Fig.~\ref{fig:reson},
we compare the scattering length computed by this method, and the
mean-field scattering length given by Eq.~(\ref{eq:TwoChannelMeanField}).
One can see from the excellent agreement that the latter is indeed
accurately computed in the stationary case. 

\section{Application: photoassociation in a sodium condensate\label{sec:results}}

\begin{figure*}
\hfill{}\includegraphics[%
  width=0.80\textwidth]{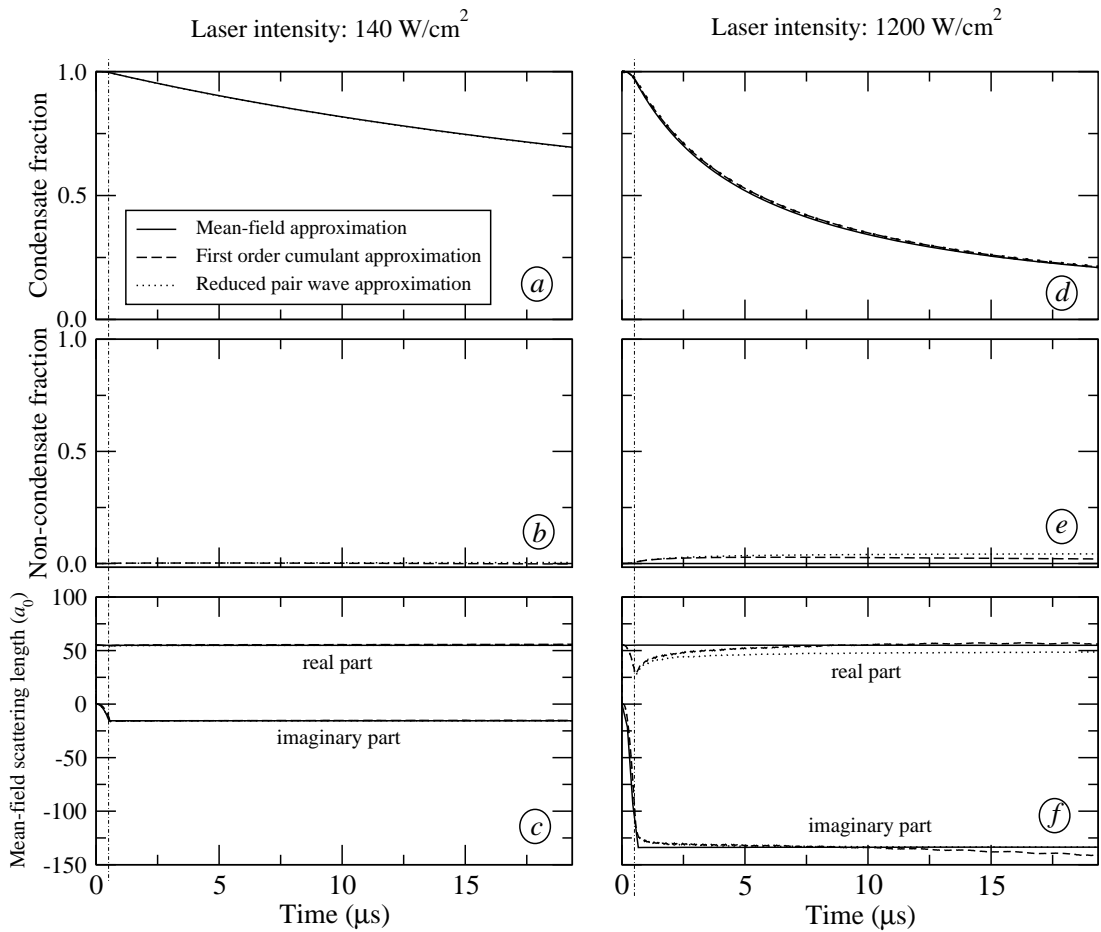}\hfill{}

\caption{On-resonance photoassociation dynamics for the $v$=135 level of
Na$_{2}$ A$^{1}\Sigma_{u}^{+}$ using a cw laser of intensity (for
$t>0.5\mu$s) $I$= 140 W/cm$^{2}$ (left column) and $I$=1200 W/cm$^{2}$
(right column). This corresponds to the adiabatic regime where all
approximations reduce to the mean-field aproximation and are consistent
with usual two-body theory. Top panels (a) and (d): the decay of the
condensate fraction $\alpha_{\text{g}}(t)$ is caused by spontaneous
emission and dissociation of the photoassociated molecules into pairs
of atoms which escape the trap. This decay is more rapid in panel
(d), due to the larger intensity (1200 W/cm$^{2}$) of the photoassociation
laser. Middle panels (b) and (e): time variation of the fraction of
non-condensate ground state atoms $\alpha_{\text{g}}^{\prime}(t)$,
which remains negligible. Bottom panels (c) and (f): time variation
of the real $\Re[a(t)]$ and imaginary $\Im[a(t)]$ part of the mean-field
scattering length.}

\label{fig:cw140}
\end{figure*}

\subsection{Experimental conditions of the NIST experiment}

\label{ssec:expcond} The chosen example starts from the photoassociation
experiment in NIST \cite{mckenzie2002}, where a condensate of sodium
atoms is illuminated by a cw laser of intensity $I$ varying from
0.14 to 1.20 kW/ cm$^{2}$. The peak density is about $4\times10^{14}$at/cm$^{3}$.
The frequency ( 16 913.37 cm$^{-1}$) is chosen at resonance with
the $J$=1, $v$=135 vibrational level in the 0$_{u}^{+}$ ($3S+3P_{1/2}$)
potential curve of Na$_{2}$. The detuning of the photoassociation
laser relative to the D$_{1}$ resonance line is rather large (43
cm$^{-1}$), and corresponds to the binding energy of the $v$=135
level. The molecules formed in the excited electronic state have a
natural width $\gamma=2\pi\times18.38$ MHz due to spontaneous emission.

Previous estimations \cite{mckenzie2002} based on criterion (\ref{eq:ConditionRogue1})
\cite{javanainen2002} predicted that these experimental conditions
should lead to rogue dissociation, as $\left(2\pi^{2}\rho\xi^{3}\right)^{-1}\approx3.2$
for $I=1.2$ kW/cm$^{2}$. However, this criterion is relevant only
in the coherent regime. According to the analysis of section \ref{sec:RogueDissociation},
the experiment is actually in the adiabatic regime (see Fig. \ref{fig:rogue}),
as $2\Omega/\gamma\sim0.04$. Therefore, the relevant criterion is
(\ref{eq:ConditionRogue2}). As $\left(2\pi^{2}\rho\xi_{0}^{3}\right)^{-1}\approx0.03$
for $I=1.2$ kW/cm$^{2}$, one concludes that no rogue dissociation
should occur and the mean-field approximation should be valid, which
is consistent with the experimental observation. However, if the intensity
can be increased up to $10$ kW/cm$^{2}$, then $\left(2\pi^{2}\rho\xi_{0}^{3}\right)^{-1}\approx0.6$
and rogue dissociation should occur (see Fig. \ref{fig:rogue}). We
therefore present numerical calculations simulating the experiment
in this high-intensity regime to see the effects of rogue dissociation.
Calculations in similar conditions were done by T.~Gasenzer \cite{gasenzer2004}
using the first-order cumulant equations with separable potentials.

\begin{figure*}
\hfill{}\includegraphics[%
  width=0.80\textwidth]{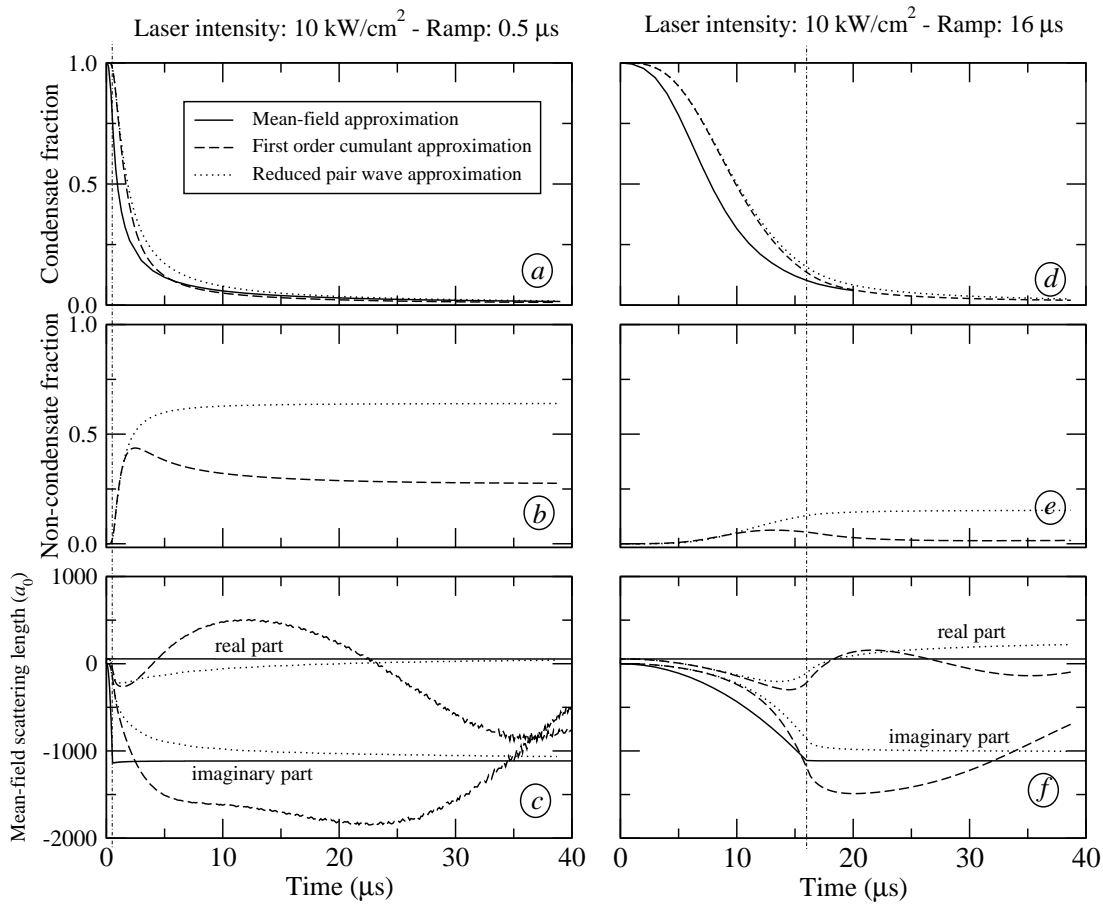}\hfill{}

\caption{Same as Fig. \ref{fig:cw140} for an intensity of 10 kW/cm$^{2}$.
Left column corresponds to a rise time of the laser of 0.5 $\mu s$,
whereas right column corresponds to a rise time of 16 $\mu s$ (the
end of the ramp is indicated by a dotted-dashed vertical line).\label{fig:cw10k}}
\end{figure*}

\subsection{Parameters of the calculations}

For these calculations, the potentials curves for the ground state
of Na$_{2}$ X$^{1}\Sigma_{g}^{+}$ and $0_{u}^{+}$(3s+3p$_{1/2})$
have been taken from Ref.~\cite{pellegrini2003}. The ground state
potential is chosen so that the scattering length is 54.9 a$_{0}$,
as in T.~Gasenzer's work \cite{gasenzer2004}. Hyperfine structure
is not included, and the last bound level has a binding energy of
317.78~MHz ( $\sim10^{-2}$ cm$^{-1}$). In the excited curve, the
level $v$=135 is bound by 49.23 cm$^{-1}$, the two neighbouring
levels being $v$=134 at -52.95~cm$^{-1}$ and $v$=136 at -45.757~cm$^{-1}$.
The spontaneous emission term (\ref{eq:imagine}) for this excited
state is set to $\gamma=2\pi\times18.36$~MHz to match the experiment
\cite{mckenzie2002}. 

The dipole moment matrix element is taken as $D$=2, assuming linear
polarisation of the laser, neglecting $r$-dependence, and deducing
the atomic lifetime from the long range coefficient $C_{3}$= 6.128
of Marinescu and Dalgarno \cite{marinescu1995}. The coupling term
$W$ in the equations (\ref{eq:2channelSchro}) is therefore linked
to the intensity I by $W=\sqrt{2I/(c\epsilon_{0})}$. 

We model the cw laser as follows: the intensity starts from zero and
is turned on linearly ($I_{b}(t)=I\times t/T;\,\, t\leq T$) to become
constant and equal to $I$ after $t=T=0.5\mu$s. This value of $T$,
previously used in the calculations by T.~Gasenzer \cite{gasenzer2004},
corresponds to the typical rise and fall-off times of the laser in
the experiment of Ref.~\cite{mckenzie2002}. Starting from a pure
sodium condensate, with a density $\rho(t=0)$=$\rho_{\text{g}}(t=0)$,
and therefore assuming $\rho^{\prime}(t=0)$=$\rho_{\text{e}}(t=0)=0$,
we have solved the coupled equations (\ref{eq:TwoChannelFirstOrder1},\ref{eq:TwoChannelFirstOrder2})
and (\ref{eq:TwoChannelReducedPair1}, \ref{eq:TwoChannelReducedPair2})
as well as the mean-field equations (\ref{eq:GrossPitaevskii1}-\ref{eq:GrossPitaevskii2}).
From these calculations, we obtain the time-variation of the relative
number of condensate atoms \begin{equation}
\alpha_{\text{g}}(t)=\rho_{\text{g}}(t)/\rho(0);\,\,\alpha_{\text{g}}(0)=1,\end{equation}
 of non-condensate atoms in the open channel, \begin{equation}
\alpha_{\text{g}}^{\prime}(t)=\rho_{\text{g}}^{\prime}(t)/\rho(0);\alpha_{\text{g}}^{\prime}(0)=0\end{equation}
 and the relative number of atoms in the closed channel \begin{equation}
\alpha_{\text{e}}(t)=\rho_{\text{e}}(t)/\rho(0)=2\beta_{\text{e}}(t);\,\,\beta_{\text{e}}(0)=0,\end{equation}
 which is twice the relative number of photoassociated molecules $\beta_{\text{e}}(t)$.
The remaining fraction $1-\alpha_{\text{g}}-\alpha_{\text{g}}^{\prime}-\alpha_{\text{e}}$
corresponds to photoassociated molecules that have been deexcited
by spontaneous emission, yielding either cold molecules in the ground
state or pairs of {}``hot atoms'' that usually leave the trap.

\subsection{Results for high intensities}

\label{ssec:results}

We first performed calculations for laser intensities $I$ in the
range of the experiment (from 140 to 1200 W/cm$^{2}$). The photoassociation
dynamics on resonance is reported in Fig.~\ref{fig:cw140}, showing
the condensate, noncondensate fractions, as well as the mean-field
scattering length as a function of time. Here we are far from the
rogue dissociation limit (see dots a and b in Fig. \ref{fig:rogue}).
As expected, the noncondensate fraction remains negligible and all
approximations agree with the mean-field approximation.

Since we are in the adiabatic regime (see Fig. \ref{fig:rogue} again),
the mean-field approximation is consistent with the usual two-body
theory. This can be seen in the left column of Fig.~(\ref{fig:lineshape}),
where we plotted the photoassociation line shape, as well as the variation
of the mean-field scattering length as a function of the frequency
of the laser. The line shape presents no difference from the one predicted
by two-body theory, nor does the mean-field scattering length from
the optically modified scattering length of the two-body theory \cite{fedichev1996,bohn1997}.
\\

We then performed calculations for $I$=10 kW/cm$^{2}$. We are now
in the rogue dissociation regime where deviations from the mean-field
approximation are expected. The on-resonance dynamics is ploted in
Fig.~\ref{fig:cw10k}. We observe the following effects in both the
FOC and RPW approximations: 

\begin{enumerate}
\item First, there is a significant final fractions of non-condensate atoms,
which of course are ignored in the Gross-Pitaevski\u{\i} picture. 
\item Second, the decay of the condensate at short times is slower than
the one predicted by the Gross-Pitaevski\u{\i} coupled equations. 
\end{enumerate}
\begin{figure}
\hfill{}\includegraphics[%
  width=0.45\textwidth]{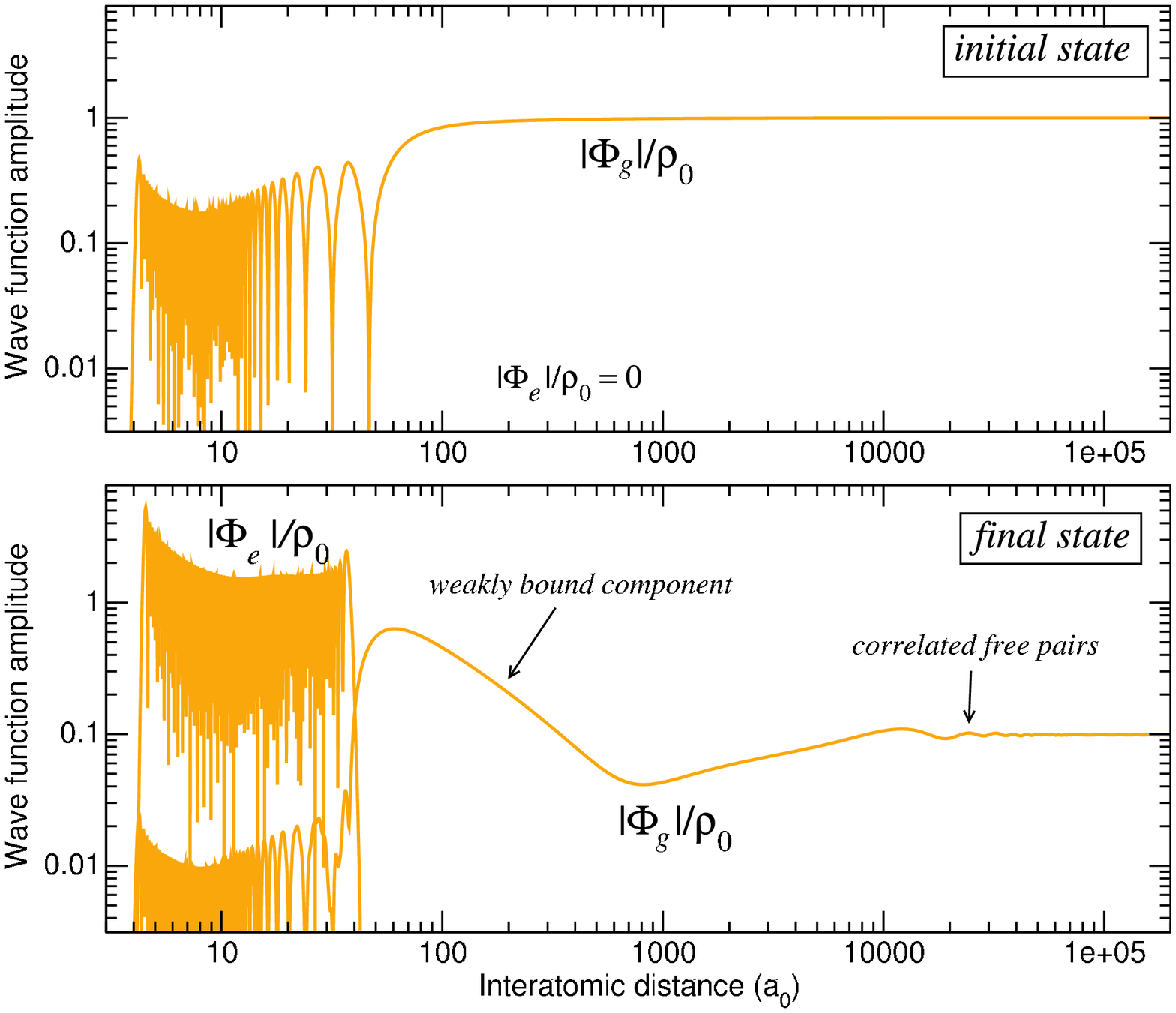}\hfill{}

\caption{Modulus of the two components $\phi_{g}(r)$ and $\phi_{e}(r)$ of
the pair wave function in the FOC approximation, as a function of
the interatomic distance. Note that a logarithmic scale is used on
both axes. Upper graph: before applying the laser: only ground state
pairs exist, $\phi_{e}(r)=0$; the wave function $\phi_{g}(r)$ corresponds
to a zero-energy scattering state: it oscillates in the short distance
region, where the potential $U_{g}(r)$ is not negligible; for distances
$r>$1000 $a_{0}$, the amplitude is equal to the initial condensate
density $\rho_{0}$. Lower graph: after applying a cw laser of intensity
$I$= 10 kW/cm$^{2}$ during 40 $\mu$s, a level of the excited potential
has been populated: $\phi_{e}(r)$ oscillates up to the classical
turning point of this level; the population transfer to this excited
state is visible in the decrease of both the short-range amplitude
of $\phi_{g}(r)$, and its very long-range amplitude corresponding
to the condensate density. Note that a strong maximum emerges around
$r\sim55$a$_{0}$; it is the near-resonance signature of a weakly
bound state in the ground-state channel. At large distances, $\phi_{g}(r)$
shows oscillations, interpreted as outward motion of {}``hot'' correlated
pairs ($T\sim$ 4 to 40 $\mu$K).}

\label{fig:comp}
\end{figure}

These two effects arise from the dynamic correlation $\Phi_{g}^{dyn}$
discussed in section \ref{sec:RogueDissociation}, but have different
interpretations.

\subsubsection{Creation of correlated pairs}

Let us first study how the non-condensate atoms are produced. This
can be done by analysing the ground state component of the pair wave
function, illustrated in Fig.~\ref{fig:comp}. We see that a strong
maximum emerges in the wave function $\Phi_{g}$ around 55~a$_{0}$,
which we identify with the presence of weakly bound molecules, corresponding
to population of the last vibrational levels in the X~$^{1}\Sigma_{g}^{+}$
potential. However, these bound states give a very small contribution
to the fraction of non-condensate atoms and they disappear when the
laser is turned off. They are in fact a near-resonance feature of
the adiabatic correlation $\Phi_{g}^{ad}$ which is already explained
by the stationary two-body theory.

The significant fraction of non-condensate atoms is explained by the
appearance of waves in the pair wave function at larger distances
(Fig.~\ref{fig:comp}). In both FOC and RPW calculation, these waves
are created at short distances as soon as the laser is turned on and
then propagate in the outward direction, reaching distances where
they are no longer affected by the laser coupling -they are not affected
when the laser is switched off. This explains why the fraction of
non-condensate atoms becomes mostly constant after 5~${\mu}$s. Thus,
these non-condensate atoms correspond to correlated pairs of free
atoms with opposite momenta. Their typical kinetic energy is expected
to be of the order of $\hbar\Gamma\sim2\,\mu K$, which is consistent
with the wave length of the waves observed in the pair wave function.
For weak trapping potentials, they may leave the system. Otherwise,
one should treat the collisions between these correlated pairs and
the condensate atoms in the presence of the laser field. However this
goes beyond both the FOC and RPW approximations. 

The appearance of the correlated pairs at high intensity has already
been predicted in T.~Gasenzer's work and it was shown that it leads
to a saturation of the number of possible stable molecules formed
by spontaneous emission \cite{gasenzer2004}. However, it is worth
noting that the formation of these pairs is a consequence of the fast
rise ($T$=0.5 $\mu$s) of the laser intensity, and is most probably
related to a two-body non-adiabatic effect - the waves also appear
in a time-dependent two-body calculation. The right column of Fig.~\ref{fig:cw10k}
shows the on-resonance dynamics when the laser intensity is rised
more slowly for $T=16$ $\mu$s. We can see that the final fraction
of non-condensate atoms is notably reduced in both approximations.
This shows that the creation of correlated pairs is greatly sensitive
to the way the laser is turned on. In contrast, the depletion of the
condensate still shows a marked deviation from the mean-field prediction.
We can then proceed with the discussion of this depletion, ignoring
the non-condensate atoms.

\subsubsection{Limitation of the decay rate}

\begin{figure*}
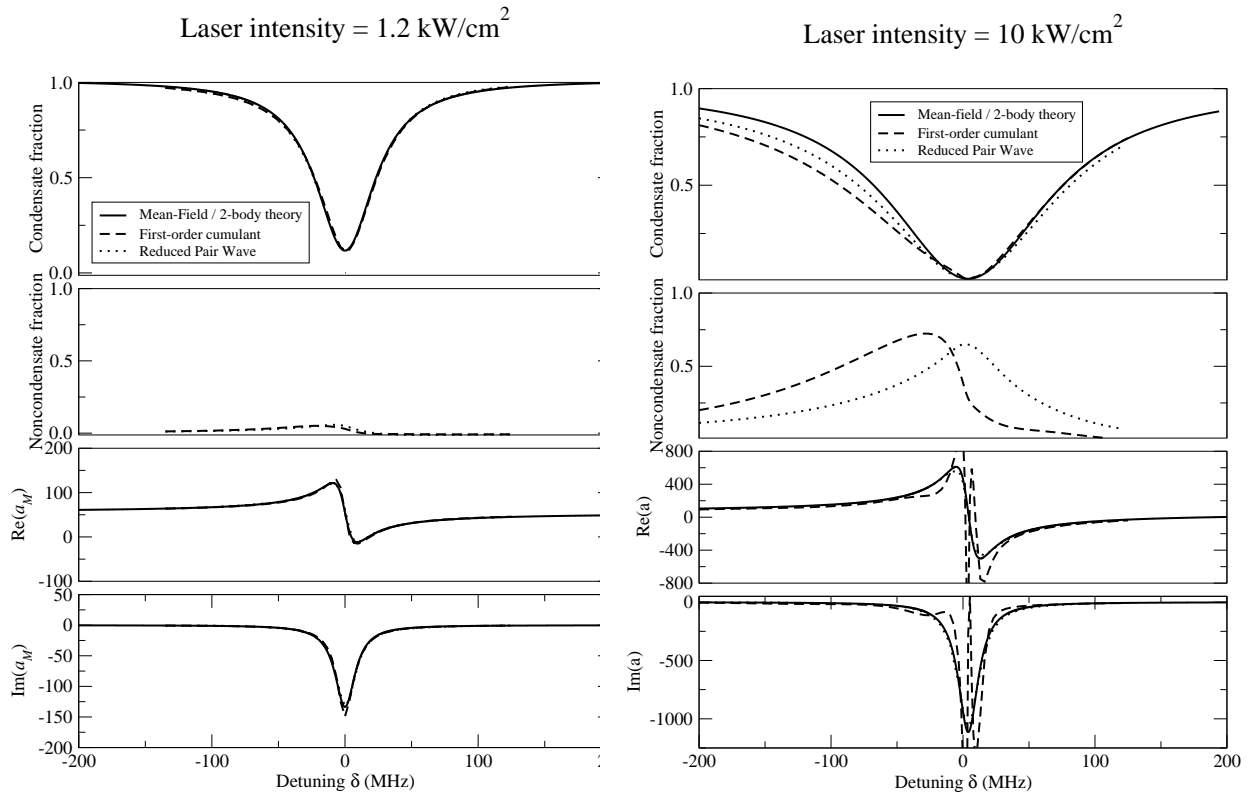

\hfill{}\includegraphics[%
  width=0.45\textwidth]{Fig8.1.eps}\includegraphics[%
  width=0.45\textwidth]{Fig8.2.eps}\hfill{}

\caption{Photoassociation line shape and optical Feshbach resonance: condensate,
non-condensate fractions and mean-field scattering length after 40~$\mu s$,
as a function of the laser detuning $\delta$ from the bare molecular
state (see Fig.~\ref{fig:pa}). At high intensity (10 kW/cm$^{2}$),
the first-order cumulant equations predict highly asymmetric behaviours
with respect to the resonance.}

\label{fig:lineshape}
\end{figure*}

In all models, the decay rate of the condensate $K(t)=\frac{h}{m}4\vert\Im{[a_{M}(t)]}\vert$
is proportional to the imaginary part of the mean-field scattering
length, the latter being one fourth of the photoassociation characteristic
length introduced in Ref.~\cite{mckenzie2002}. The imaginary part
of the mean-field scattering is plotted in the bottom panels of Fig.~\ref{fig:cw10k}.
At short times, both the FOC and RPW approximations lead to the same
rate, which is clearly smaller than the one predicted by the mean-field
approximation. Such a limitation of the rate was already pointed out
in the coherent regime \cite{javanainen2002}, and also observed in
the adiabatic regime \cite{gasenzer2004}.

Our interpretation is the following: because of the rogue dissociation,
excited molecules are coupled back to the condensate through the dynamic
correlation. This extra correlation reduces the efficiency of the
coupling between the condensate and the excited molecules, thereby
reducing the loss from the condensate. The fact that this effect does
not depend on the way the laser is turned on suggests that it is a
genuine many-body effect, namely the influence of the dynamics of
the medium on the pair dynamics. We also note that the dynamic correlation
responsible for this limitation of the rate gives a very small contribution
to the non-condensate atoms, unlike the part corresponding to the
correlated free atoms. We suspect from Eqs. (\ref{eq:GrossPitaevskiiDyn1}-\ref{eq:dynRPW})
that the Fourier transform of this correlation scales as $1/k^{4}$.

Note that at longer times in the reduced pair wave approximation,
the rate (and more generally the mean-field scattering length) approaches
the standard two-body rate (standard scattering length $A$). On the
other hand, in the first-order cumulant approximation the rate finally
exceeds the standard rate and starts oscillating. However, these differences
are not so significant from the experimental point of view because
they occur at a point where there are very few condensate atoms remaining
in the system.

\subsubsection{Symmetry of the line shape}

The differences between the RPW anf FOC approximations are more conspicuous
in the line shape of the resonance - see right column of Fig. \ref{fig:lineshape}.
In the RPW approximation the final condensate and non-condensate fractions
are symmetric with respect to the resonance. On the other hand, in
the FOC approximation the line shape becomes very asymmetric: on the
{}``blue'' side of the resonance (where the bare resonant state
lies below the dressed threshold) the condensate is more depleted
and more non-condensate atoms are produced than on the {}``red''
side. We suspect that this asymmetry is due to the strong dependence
of the scattering properties on the mean-field energy in the FOC approximation,
as we saw in Section \ref{sub:FOC-stationary-states}. Indeed, at
high-intensity the mean-field energy becomes large and positive on
the blue side while it is large and negative on the red side. On the
other hand, the scattering properties in the RPW approximation correspond
mainly to those of the usual zero energy scattering problem.

Comparison of these predictions with experiment may not be straightforward.
First, there are important experimental issues such as inhomogeneous
broadenening to overcome at these high intensities \cite{mckenzie2002}.
Second, collisions between condensate and non-condensate atoms are
neglected in both approximations, and as we already pointed out in
Section \ref{sub:Reduced-pair-wave}, the RPW approximation brings
some correction to the FOC approximation, but only in an incomplete
way. However, at a qualitative level, the symmetry of the experimental
line shapes at these high intensities could show the relevance of
this mean-field correction occurring at short interatomic distance.

\subsubsection{Realistic potentials and cw lasers}

Finally, we should note that our calculations using the FOC approximation
with realistic potentials agree with T.~Gasenzer's calculations which
use a separable potential \cite{gasenzer2004}, except for times when
the condensate fraction becomes negligible. This shows that no major
improvement is brought by the details of the potential in these cw
laser photoassociation calculations. Using the simplified equations
(\ref{eq:GrossPitaevskiiDyn1}-\ref{eq:dynRPW}) in this case would
lead to similar results. In fact, since we have addressed only the
adiabatic regime (see Fig.~\ref{fig:rogue}), only two equations
would be sufficient, namely (\ref{eq:GrossPitaevskiiDyn1}) and (\ref{eq:dynFOC})
or (\ref{eq:GrossPitaevskiiDyn1}) and (\ref{eq:dynRPW}), as the
molecular condensate $\Psi_{m}$can be eliminated adiabatically in
these equations. Note that these equations are very similar to those
of say, Ref.~\cite{javanainen2002}, but are free of any ultraviolet
divergence or renormalization process as the adiabatic correlation
have been eliminated from the equations in the first place.

In the more complex photoassociation processes involving pulsed lasers,
the simplified equations are inadequate and we shall use the more
general equations and numerical procedures described in this paper
to investigate these situations.

\section{Conclusion}

\label{sec:conclu} We have compared two many-body approximations
for the time-dependent description of photoassociation and optically-induced
Feshbach resonances in an atomic condensate: the first-order cumulant
approximation and the reduced pair wave approximation. These approximations
differ only in the way the influence of the mean field on a pair of
condensate atoms is treated at short separations. Each approximation
leads to a set of coupled equations describing the microscopic and
macroscopic dynamics for a two-channel problem. We demonstrated that
these equations can be solved numerically with realistic molecular
potentials, which should prove essential when addressing experiments
with chirped laser pulses, for which details of the interaction potentials
may influence the dynamics.

From the general equations, we identified several regimes. We define
the adiabatic regime when the excited molecular channel is just an
intermediate state during the collision of two condensate atoms. In
contrast, we define the coherent regime when the excited channel gives
rise to a molecular condensate having the features of a one-body condensate.
In each of these two regimes, a mean-field theory is obtained when
the pair correlation in the ground channel can be eliminated adiabatically.
In the coherent regime, this leads to two coupled Gross-Pitaevski\u{\i}
equations \cite{timmermans1999}. In the adiabatic regime, this leads
to a single Gross-Pitaevski\u{\i} equation with a complex scattering
length predicted by two-body theories. This is the usual regime investigated
so far experimentally \cite{mckenzie2002,theis2004,winkler2005}.

The condition for the breakdown of the mean-field approximation (or
so-called rogue dissociation \cite{javanainen2002}) is different
in each regime. We showed that, contrary to previous estimates \cite{javanainen2002,mckenzie2002},
the condition for current experiments are under the adiabatic regime. 

Solving the general equations numerically, we investigated the case
of a sodium condensate where a photoassociation cw laser is turned
on in conditions similar to the experiment of McKenzie \textit{et
al} \cite{mckenzie2002}. At high intensities ($\sim$10 kW/cm$^{2}$),
which could not be attained in \cite{mckenzie2002} and where rogue
dissociation is expected to occur, we observed the following effects: 

\begin{enumerate}
\item Ground-state correlated pairs of free atoms are produced. This is
agrees with the saturation predicted by T.~Gasenzer \cite{gasenzer2004}.
The creation of such atom pairs can indeed strongly limit the final
yield of ground-state molecules formed by spontaneous decay. However,
Ref.~\cite{gasenzer2004} considered only a rapid turn on of the
laser. By introducing a slower switching procedure, leading to a more
adiabatic behaviour, the present work shows that the production of
hot atom pairs may be decreased. 
\item The photoassociation rate of the condensate at short times is smaller
than the rate predicted in the mean-field approximation. This limitation
happens at much lower intensities than the unitary limit of the two-body
theory. This effect has already been predicted by J.~Javanainen and
M.~Mackie \cite{javanainen2002} in the coherent regime. We interpret
it as the appearance of a dynamic many-body correlation in the condensate.
\item The photoassociation line shape becomes asymmetric in the first-order
cumulant approximation, while it remains symmetric in the reduced
pair wave approximation. 
\end{enumerate}
We think this last point should be a good qualitative effect to experimentally
distinguish between the two approximations. Future work will investigate
the time dependence of the formation of stable molecules, via two-colour
or one-colour experiments with pulsed lasers, using the theoretical
and numerical tools developed in the present paper. \\
\\

\textbf{\underbar{Acknowledgements}} P.N. is very grateful to Eite
Tiesinga for his essential remarks on this work. Discussions with
Arkady Shanenko, Paul Julienne, Thorsten K\"{o}hler, Christiane Koch,
Eliane Luc-Koenig, Philippe Pellegrini are gratefully acknowledged
as well. This work was performed in the framework of the European
Research Training Network {}``Cold Molecules'', funded by the European
Commission under contract HPRN CT 2002 00290. P.N. acknowledges for
an invitation in Clarendon Laboratory, Oxford. 

\newpage

\section{Appendix}

\subsection{Definition of the non-commutative cumulants}

If $\hat{A}_{1},\;\hat{A}_{2},\:...\hat{A}_{n}$ are $n$ bosonic
operators, the $n$-order cumulant $\langle\hat{A}_{1}\hat{A}_{2}...\hat{A}_{n}\rangle^{c}$
is defined recursively as follows~\cite{kohler2002}: 

\begin{eqnarray*}
\langle\hat{A}_{1}\rangle & = & \langle\hat{A}_{1}\rangle^{c}\\
\langle\hat{A}_{1}\hat{A}_{2}\rangle & = & \langle\hat{A}_{1}\rangle^{c}\langle\hat{A}_{2}\rangle^{c}+\langle\hat{A}_{1}\hat{A}_{2}\rangle^{c}\\
\langle\hat{A}_{1}\hat{A}_{2}\hat{A}_{3}\rangle & = & \langle\hat{A}_{1}\rangle^{c}\langle\hat{A}_{2}\rangle^{c}\langle\hat{A}_{3}\rangle^{c}\;+\;\langle\hat{A}_{1}\rangle^{c}\langle\hat{A}_{2}\hat{A}_{3}\rangle^{c}\\
 &  & +\;\langle\hat{A}_{2}\rangle^{c}\langle\hat{A}_{1}\hat{A}_{3}\rangle^{c}\;+\;\langle\hat{A}_{3}\rangle^{c}\langle\hat{A}_{1}\hat{A}_{2}\rangle^{c}\\
 &  & +\;\langle\hat{A}_{1}\hat{A}_{2}\hat{A}_{3}\rangle^{c}\\
\; & \vdots & \;\end{eqnarray*}
where $\langle...\rangle$ denotes the quantum average in the many-body
states of the system. For an ideal gas, we can check that all cumulants
containing more than two bosonic operators are zero, according to
Wick's theorem. For a system close to the ideal gas, the cumulants
are non-zero but tend to zero as the order is increased. They are
therefore a sort of measure of the deviation from the ideal case.

\subsection{In-medium effective wave functions}

\subsubsection{General expressions}

The in-medium pair wave function approach, as we may call it, was
initiated by a series of papers by A. Yu. Cherny and A. A. Shanenko.
Its aim is to have a many-body description of the dilute Bose gas
which remains valid at short interatomic distances, so that interaction
potentials with strong repulsive cores can be treated directly. The
authors have mainly addressed the problem of the ground state for
a homogeneous system. In Ref.~\cite{naidon2003}, we have generalized
some of their ideas to the inhomogeneous time-dependent case. We will
recall here the derivation of the in-medium pair wave functions given
in \cite{naidon2003}, and will show in addition how three-body wave
functions can be constructed out of these pair wave functions. This
will lead to the reduced pair wave approximation used in this paper. 

The starting point is to consider the reduced density matrices of
the the many-boson system. As these matrices are hermitian, they can
be diagonalized in a basis of orthogonal eigenvectors, associated
with positive eigenvalues. For example, the following one-body, and
two-body reduced density matrices can be diagonalized as follows:\begin{eqnarray}
\langle\hat{\psi}^{\dagger}(\mathbf{x})\hat{\psi}(\mathbf{y})\rangle & = & \sum_{i}\Psi_{i}^{*}(\mathbf{x})\Psi_{i}(\mathbf{y})\label{eq:DiagOneBodyDensity}\\
\frac{1}{2}\langle\hat{\psi}^{\dagger}(\mathbf{w})\hat{\psi}^{\dagger}(\mathbf{z})\hat{\psi}(\mathbf{x})\hat{\psi}(\mathbf{y})\rangle & = & \sum_{i}\Phi_{i}^{*}(\mathbf{w,z})\Phi_{i}(\mathbf{x,y})\qquad\label{eq:DiagTwoBodyDensity}\end{eqnarray}
where the $\Psi_{i}$ are the eigenvectors of the one-body density
matrix, and $\Phi_{i}$ are the eigenvectors of the two-body density
matrix. We have normalized these vectors to their respective eigenvalues,
which means that $N_{i}=\int d^{3}\mathbf{x}|\Psi_{i}(\mathbf{x})|^{2}$
is the eigenvalue associated with $\Psi_{i}$, and $M_{i}=\int d^{3}\mathbf{y}d^{3}\mathbf{x}|\Phi_{i}(\mathbf{x,y})|^{2}$
is the eigenvalue associated with $\Phi_{i}$. As these eigenvectors
have the form of wave functions, we call them effective one-body wave
functions and effective pair wave functions. $N_{i}$ is interpreted
as the average number of particles of the system in the one-body state
$|\Psi_{i}\rangle$, and $M_{i}$ is interpreted as the average number
of pairs in the two-body state $|\Phi_{i}\rangle$. We can check that
$\sum_{i}N_{i}=N$ and $\sum_{i}M_{i}=\frac{1}{2}N(N-1)$, where $N$
is the total number of particles in the system. 

In the case of a condensate, a certain one-body wave function $\Psi_{0}$
has a macroscopic occupation number $N_{0}\gg\sum_{i\ne0}N_{i}$.
In the $U(1)$ symmetry breaking picture, this one-body wave function
is identified with the condensate wave function, or order parameter
$\langle\hat{\psi}\rangle$. According to the Bogoliubov prescription,
the field operator $\hat{\psi}$ can then be decomposed into its average
value $\Psi_{0}$ and a remaining fluctuating operator $\hat{\theta}$.
Expanding the two-body density matrix (\ref{eq:DiagTwoBodyDensity})
with the Bogoliubov prescription, and refactorising the expression,
we showed that it can be written:\begin{eqnarray}
 &  & \sum_{i=0}^{\infty}\Phi_{0i}^{*}(\mathbf{w,z})\Phi_{0i}(\mathbf{x,y})\label{eq:Factorization1}\\
 &  & +\frac{1}{2}\langle\hat{\theta}^{\dagger}(\mathbf{w})\hat{\theta}^{\dagger}(\mathbf{z})\hat{\theta}(\mathbf{x})\hat{\theta}(\mathbf{y})\rangle-\sum_{i\ne0}^{\infty}{\Phi'}_{0i}^{*}(\mathbf{w,z})\Phi'_{0i}(\mathbf{x,y})\nonumber \end{eqnarray}
where:\begin{eqnarray}
\Phi_{00}(\mathbf{x,y}) & = & \frac{1}{\sqrt{2}}\Psi_{0}(\mathbf{x})\Psi_{0}(\mathbf{y})\;+\;\Phi_{00}^{\prime}(\mathbf{x,y})\label{eq:PairWave1}\\
\Phi_{0i}(\mathbf{x,y}) & = & \frac{\Psi_{0}(\mathbf{x})\Psi_{i}(\mathbf{y})+\Psi_{0}(\mathbf{y})\Psi_{i}(\mathbf{x})}{\sqrt{2}}+\Phi_{0i}^{\prime}(\mathbf{x,y})\qquad\label{eq:PairWave2}\end{eqnarray}

for $i\ne0$ and:\begin{eqnarray*}
\Phi_{00}^{\prime}(\mathbf{x,y}) & = & \frac{1}{\sqrt{2}}\langle\hat{\theta}(\mathbf{x})\hat{\theta}(\mathbf{x})\rangle\\
\Phi_{0i}^{\prime}(\mathbf{x,y}) & = & \int d^{3}\mathbf{z}\frac{\Psi_{i}(\mathbf{z})}{\sqrt{2N_{i}}}\langle\hat{\theta}^{\dagger}(\mathbf{z})\hat{\theta}(\mathbf{x})\hat{\theta}(\mathbf{y})\rangle\quad\text{for $i\ne0$}\end{eqnarray*}

As the $\Phi_{0i}$'s and $\Phi_{00}$ are orthogonal in the limit
of large systems, we conclude that we have found the first pair wave
functions of the diagonalized form (\ref{eq:DiagTwoBodyDensity}). 

$\Phi_{00}$ is interpreted as the condensate-condensate wave function,
\emph{ie} the wave function for a pair of particles both coming from
the condensed part. It is composed of two terms: an asymptotic term
$\frac{1}{\sqrt{2}}\Psi_{0}(\mathbf{x})\Psi_{0}(\mathbf{y})$ corresponding
to the free motion of two independent condensate particles, and a
scattering term $\Phi_{00}^{\prime}$ corresponding to their correlated
motion due to their interaction. Similarly, $\Phi_{0i}$ for $i\ne0$
is interpreted as the condensate-noncondensate wave function, \emph{ie}
the wave function for two particles, one coming from the condensed
part and the other coming from the non-condensed part. It is also
composed of an asymptotic term $\frac{\Psi_{0}(\mathbf{x})\Psi_{i}(\mathbf{y})+\Psi_{0}(\mathbf{y})\Psi_{i}(\mathbf{x})}{\sqrt{2}}$
(note the symmetrization) and a scattering term $\Phi_{0i}^{\prime}$.
We can check that for a sufficiently large system the norm of $\Phi_{00}$
is $\frac{1}{2}N_{0}^{2}$, the number of condensate pairs, and the
norm of $\Phi_{0i}$ is $N_{0}N_{i}$, the number of pairs involving
a condensate particle and a non-condensate particle in state $|\Psi_{i}\rangle$.
The presence of a condensate therefore implies a condensation of unbound
pairs, as $\frac{1}{2}N_{0}^{2}\gg\sum_{i\ne0}N_{0}N_{i}$. 

The last line of Eq.~(\ref{eq:Factorization1}) corresponds to the
remaining pair wave functions of the system. We have no explicit expression
for those, but if we assume that the non-condensate particles form
an ideal gas (\emph{ie} do not interact), we can use Wick's theorem
to express the last line, and check that it is equal to:\[
\sum_{0<i\leq j}\Phi_{ij}^{*}(\mathbf{w,z})\Phi_{ij}(\mathbf{x,y})\]
with:\begin{eqnarray*}
\Phi_{ij}(\mathbf{x,y}) & = & \Psi_{i}(\mathbf{x})\Psi_{i}(\mathbf{y})\\
\Phi_{ij}(\mathbf{x,y}) & = & \frac{\Psi_{i}(\mathbf{x})\Psi_{j}(\mathbf{y})+\Psi_{i}(\mathbf{y})\Psi_{j}(\mathbf{x})}{\sqrt{2}}\quad\text{for }i\ne j\end{eqnarray*}
 which are indeed the expected pair wave functions for two non-interacting
non-condensate atoms %
\footnote{However, Wick's theorem, like the Hartree-Fock decoupling, induces
a double-counting of non-condensate particles in the same state which
we have missed in Ref.~\cite{naidon2003}. Eq.~(A23) of this reference
should not contain the factor $\sqrt{2}$. Note however that this
is an artifact of the Wick/Hartree-Fock decoupling and that the factor
$\sqrt{2}$ is really present for a non-interacting system, as can
be checked by direct calculation.%
}. 

For convenience, we may rewrite the scattering terms appearing in
(\ref{eq:PairWaveCondensate}) and (\ref{eq:PairWaveNonCondensed})
with multiplicative correlation functions $\varphi_{0i}$ defined
as follows:\begin{eqnarray}
\Phi_{00}(\mathbf{x,y}) & \equiv & \frac{1}{\sqrt{2}}\Psi_{0}(\mathbf{x})\Psi_{0}(\mathbf{y})\varphi_{00}(\mathbf{x,y})\label{eq:PairWaveCondensate}\\
\Phi_{0i}(\mathbf{x,y}) & \equiv & \frac{\Psi_{0}(\mathbf{x})\Psi_{i}(\mathbf{y})+\Psi_{0}(\mathbf{y})\Psi_{i}(\mathbf{x})}{\sqrt{2}}\varphi_{0i}(\mathbf{x,y})\qquad\label{eq:PairWaveNonCondensed}\end{eqnarray}

\begin{figure}
\hfill{}\includegraphics[%
  width=0.45\textwidth]{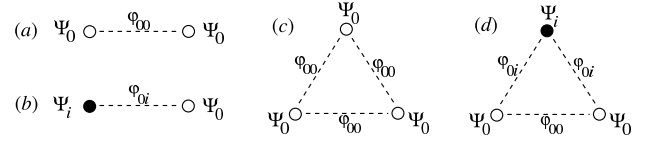}\hfill{}

\caption{{\small Schematic reprensentations of some effective in-medium wave
functions in a condensate: the condensate pair wave function $(a)$,
the condensate-noncondensate pair wave function $(b)$, the condensate
three-body wave function $(c)$, and the condensate-condensate-noncondensate
three-body wave function $(d)$. The circles indicate the asymptotic
one-body behaviours, and the dashed lines represent the correlation.}}

\label{Fig:Diagrams}
\end{figure}

We call these functions $\varphi_{0i}$ the reduced pair wave functions.
Assuming that the particles must be decorrelated at large distances,
we expect that the reduced pair wave functionstend to 1 at large distances.
Thus, we may write $\varphi_{0i}=1+\varphi_{0i}^{\prime}$ for convenience,
with $\varphi_{0i}^{\prime}$ vanishing at large distances. Another
important property is that for an interaction with a strong repulsive
core, the probability of finding two particles close to each other
should tend to zero at short distances; we therefore expect that the
reduced pair wave functions$\varphi_{0i}$ tend to zero at short distances.
This is precisely because they tend to zero at short distances that
they have a regularising effect when multiplied with the interaction
potential. 

We can now express the quantum averages found in Eqs.~(\ref{eq:ExactCondensateEquation})
and (\ref{eq:ExactPairEquation}) in terms of pair wave functions.
By expanding the quantum average in Eq.~(\ref{eq:ExactCondensateEquation})
with the Bogoliubov prescription and refactorising the terms, we find:\begin{equation}
\langle\hat{\psi}^{\dagger}(\textbf{z})\hat{\psi}(\textbf{z})\hat{\psi}(\mathbf{x})\rangle=\sqrt{2}\sum_{i=0}^{\infty}\Psi_{i}^{*}(\mathbf{z})\Phi_{0i}(\mathbf{z,x})\label{eq:Factorization2}\end{equation}

This gives a clear interpretation of this term in Eq.~(\ref{eq:ExactCondensateEquation}):
the condensate particles can collide either with another condensate
particle or with a non-condensate particle. Each type of collision
gives rise to a term looking like a scattering amplitude, where the
corresponding two-body wave function is involved. 

Similarly, by expanding the quantum average in Eq.~(\ref{eq:ExactPairEquation})
in terms of cumulants, and refactorising the terms, we find:\begin{equation}
\langle\hat{\psi}^{\dagger}(\textbf{z})\hat{\psi}(\textbf{z})\hat{\psi}(\mathbf{x})\hat{\psi}(\mathbf{y})\rangle=\sqrt{3!}\sum_{i=0}^{\infty}\Psi_{i}^{*}(\mathbf{z})\Omega_{00i}(\mathbf{z,x,y})\label{eq:Factorization3}\end{equation}
with the three-body wave functions: 

\begin{widetext}
\begin{eqnarray*}
\Omega_{000}(\mathbf{z,x,y}) & \equiv & \;\frac{1}{\sqrt{3!}}\Psi_{0}(\!\mathbf{z}\!)\Psi_{0}(\!\mathbf{x}\!)\Psi_{0}(\!\mathbf{y}\!)\:\Big(1+{\varphi'}_{00}(\mathbf{z,\! x})+{\varphi'}_{00}(\mathbf{z,\! y})+{\varphi'}_{00}(\mathbf{x,\! y})\Big)\;+\;\Omega_{000}^{\prime}(\mathbf{z,x,y})\\
\\\Omega_{00i}(\mathbf{z,x,y}) & \equiv & \;\frac{1}{\sqrt{3!}}\left(\begin{array}{c}
\quad\Psi_{i}(\!\mathbf{z}\!)\Psi_{0}(\!\mathbf{x}\!)\Psi_{0}(\mathbf{y})\:\Big(1+\varphi_{0i}^{\prime}(\mathbf{z,\! x})+\varphi_{0i}^{\prime}(\mathbf{z,\! y})+\varphi_{00}^{\prime}(\mathbf{x,\! y})\Big)\\
+\;\Psi_{i}(\!\mathbf{y}\!)\Psi_{0}(\!\mathbf{z}\!)\Psi_{0}(\mathbf{x})\:\Big(1+\varphi_{00}^{\prime}(\mathbf{z,\! x})+\varphi_{0i}^{\prime}(\mathbf{z,\! y})+\varphi_{0i}^{\prime}(\mathbf{x,\! y})\Big)\\
+\;\Psi_{i}(\!\mathbf{x}\!)\Psi_{0}(\!\mathbf{z}\!)\Psi_{0}(\mathbf{y})\:\Big(1+\varphi_{0i}^{\prime}(\mathbf{z,\! x})+\varphi_{00}^{\prime}(\mathbf{z,\! y})+\varphi_{0i}^{\prime}(\mathbf{x,\! y})\Big)\end{array}\right)\;+\;\Omega_{00i}^{\prime}(\mathbf{z,x,y})
\end{eqnarray*}
\end{widetext}

where:\begin{eqnarray*}
\Omega'_{000}(\mathbf{z,x,y}) & = & \frac{1}{\sqrt{3!}}\langle\hat{\theta}(\mathbf{z})\hat{\theta}(\mathbf{x})\hat{\theta}(\mathbf{y})\rangle^{c}\\
\Omega'_{00i}(\mathbf{z,x,y}) & = & \int\!\! d^{3}\mathbf{w}\frac{\Psi_{i}(\mathbf{w})}{\sqrt{3!}N_{i}}\langle\hat{\theta}^{\dagger}(\mathbf{w})\hat{\theta}(\mathbf{z})\hat{\theta}(\mathbf{x})\hat{\theta}(\mathbf{y})\rangle^{c}\end{eqnarray*}

We note that the first term of each three-body wave functions is not
the wave function of independent particles, contrary to what we found
with pair wave functions. Indeed, it already contains some pair correlation
through the terms $\varphi'$. However, these pair correlations are
not complete. For an interaction with a strong repulsive core, the
probability for finding three atoms very close to one another must
tend to zero. But here, the first term of each three-body wave function
does not vanish when $\mathbf{x}\approx\mathbf{y}\approx\mathbf{z}$.
We think this is because this first term is only the asymptotic behaviour
of the three-body wave function at large distances. The terms $\Omega^{\prime}$
must contain stronger correlations at short distances. We expect that
at short distances, the three-body wave functions are in fact fully
correlated through a product of reduced pair wave functions akin to
a Jastrow wave function\cite{jastrow1955}: 

\begin{widetext}
\begin{eqnarray*}
\Omega_{000}(\mathbf{z,x,y}) & \equiv & \;\frac{1}{\sqrt{3!}}\Psi_{0}(\!\mathbf{z}\!)\Psi_{0}(\!\mathbf{x}\!)\Psi_{0}(\!\mathbf{y}\!)\:{\varphi}_{00}(\mathbf{z,\! x}){\varphi}_{00}(\mathbf{z,\! y}){\varphi}_{00}(\mathbf{x,\! y})\;+\;\Omega_{000}^{\prime\prime}(\mathbf{z,x,y})\\
\\\Omega_{00i}(\mathbf{z,x,y}) & \equiv & \;\frac{1}{\sqrt{3!}}\left(\begin{array}{c}
\quad\Psi_{i}(\!\mathbf{z}\!)\Psi_{0}(\!\mathbf{x}\!)\Psi_{0}(\mathbf{y})\:{\varphi}_{0i}(\mathbf{z,\! x}){\varphi}_{0i}(\mathbf{z,\! y}){\varphi}_{00}(\mathbf{x,\! y})\\
+\;\Psi_{i}(\!\mathbf{y}\!)\Psi_{0}(\!\mathbf{z}\!)\Psi_{0}(\mathbf{x})\:{\varphi}_{00}(\mathbf{z,\! x}){\varphi}_{0i}(\mathbf{z,\! y}){\varphi}_{0i}(\mathbf{x,\! y})\\
+\;\Psi_{i}(\!\mathbf{x}\!)\Psi_{0}(\!\mathbf{z}\!)\Psi_{0}(\mathbf{y})\:{\varphi}_{0i}(\mathbf{z,\! x}){\varphi}_{00}(\mathbf{z,\! y}){\varphi}_{0i}(\mathbf{x,\! y})\end{array}\right)\;+\;\Omega_{00i}^{\prime\prime}(\mathbf{z,x,y})\end{eqnarray*}
\end{widetext}

This is the most natural structure which preserves the symmetry of
the three-body wave functions and leads to a zero probability at short
distances. The remaining terms $\Omega^{\prime\prime}$ are supposed
to contain the three-body correlations which cannot be expressed in
terms of two-body correlations. Note again that the norm of $\Omega_{000}$
is $\frac{1}{3!}N_{0}^{3}$ corresponding to the number of condensate
triplets, while the norm of $\Omega_{00i}$ is $\frac{1}{2}N_{0}^{2}N_{i}$,
corresponding to the number of triplets involving two condensate particles
and one particle in the non-condensate state $|\Psi_{i}\rangle$. 

The interpretation of (\ref{eq:Factorization3}) appearing in Eq.~(\ref{eq:ExactPairEquation})
is again very simple: the pairs of condensate atoms can collide either
with another condensate atom, or with a non-condensate atom. Each
type of collision is accounted for by a scattering amplitude-like
term, involving the corresponding three-body wave function.

\subsubsection{Simplified expressions}

Neglecting the collisions with non-condensate particles as well as
the three-body correlation $\Omega_{000}^{\prime\prime}$, we can
express the quantum averages in terms of only the condensate wave
function $\Psi\equiv\Psi_{0}$ and the condensate pair wave function
$\Phi=\sqrt{2}\Phi_{00}$:\begin{eqnarray}
\langle\hat{\psi}^{\dagger}(\textbf{w})\hat{\psi}(\textbf{z})\hat{\psi}(\mathbf{x})\rangle & \approx & \Psi^{*}(\mathbf{w})\Phi(\mathbf{z,x})\label{eq:SimplifiedExpr1}\\
\langle\hat{\psi}^{\dagger}(\mathbf{w})\hat{\psi}^{\dagger}(\mathbf{z})\hat{\psi}(\mathbf{x})\hat{\psi}(\mathbf{y})\rangle & \approx & \Phi^{*}(\mathbf{w,z})\Phi(\mathbf{x,y})\label{eq:SimplifiedExpr2}\\
\langle\hat{\psi}^{\dagger}(\textbf{w})\hat{\psi}(\textbf{z})\hat{\psi}(\mathbf{x})\hat{\psi}(\mathbf{y})\rangle & \approx & \Psi^{*}(\mathbf{w})\frac{\Phi(\mathbf{z,x})\Phi(\mathbf{z,y})\Phi(\mathbf{x,y})}{\Psi(\mathbf{x})\Psi(\mathbf{y})\Psi(\mathbf{z})}\qquad\label{eq:SimplifiedExpr3}\end{eqnarray}

Note that all these expressions respect the symmetry of the quantum
averages by exchange of coordinates. Approximations (\ref{eq:SimplifiedExpr1})
and (\ref{eq:SimplifiedExpr2}) are found within the first-order cumulant
approach, but not Approximation (\ref{eq:SimplifiedExpr3}). The reduced
pair wave approximation (\ref{eq:ReducedPairApproximation}) used
in this paper is a simplified version of (\ref{eq:SimplifiedExpr3})
where the correlation between $\mathbf{z}$ and $\mathbf{x}$ is neglected. 

\bibliographystyle{apsrev}
\bibliography{biblio}

\end{document}